\PassOptionsToPackage{rgb,table,xcdraw}{xcolor}
\PassOptionsToPackage{left=1cm,right=1.5cm,bottom=2cm}{geometry}

\documentclass[sn-nature]{sn-jnl}

\usepackage[fleqn]{amsmath}
\usepackage{amssymb}
\usepackage{amsfonts}
\usepackage{bm}

\usepackage{graphicx}
\usepackage{url}
\usepackage{color}
\usepackage{mathtools}
\usepackage{scalerel}
\usepackage{subfig}
\usepackage{ragged2e}
\usepackage{wrapfig}
\usepackage[misc]{ifsym}

\usepackage{breakcites}
\usepackage{microtype}

\usepackage{tikz}
\usetikzlibrary{arrows}
\usetikzlibrary{automata}
\usetikzlibrary{calc}
\usetikzlibrary{shapes}
\usetikzlibrary{positioning}

\usepackage{pgfplots}
\pgfplotsset{compat=1.18}

\usepackage{pgfplotstable}
\pgfplotstableset{
    color cells/.style={
        col sep=comma,
        string type,
        postproc cell content/.code={%
                \pgfkeysalso{@cell content=\rule{0cm}{2.4ex}\cellcolor{black!##1}}%
                },
        columns/x/.style={
            column name={},
            postproc cell content/.code={}
        }
    }
}

\usepackage{algorithm}
\usepackage{algpseudocode}

\usepackage{orcidlink}
\usepackage[numbers,sort&compress]{natbib}

\usepackage[shortlabels]{enumitem}

\usepackage{booktabs}\let\cline\cmidrule % Fix '\cline' not working.

\begin{document}

\title{Overview and Performance Evaluation of Supervisory Controller Synthesis with Eclipse ESCET v4.0}

\author*[1,2]{\fnm{Dennis} \sur{Hendriks} \orcidlink{0000-0002-9886-7918}}
\email{dennis.hendriks@tno.nl}\email{dennis.hendriks@ru.nl}

\author[3]{\fnm{Michel} \sur{Reniers} \orcidlink{0000-0002-9283-4074}}
\email{m.a.reniers@tue.nl}

\author[3,4]{\fnm{Wan} \sur{Fokkink} \orcidlink{0000-0001-7443-8978}}
\email{w.j.fokkink@vu.nl}

\author[1]{\fnm{Wytse} \sur{Oortwijn} \orcidlink{0000-0002-5244-2519}}
\email{wytse.oortwijn@tno.nl}

\affil*[1]{
    \orgname{TNO-ESI},
    \orgaddress{\street{High Tech Campus 25}, \postcode{5656 AE}, \city{Eindhoven}, \country{The Netherlands}}%
}
\affil[2]{
    \orgname{Radboud University},
    \orgaddress{\street{Toernooiveld 212}, \postcode{6525 EC}, \city{Nijmegen}, \country{The Netherlands}}%
}
\affil[3]{
    \orgname{Eindhoven University of Technology},
    \orgaddress{\street{Het Eeuwsel 2}, \postcode{5612 AS}, \city{Eindhoven}, \country{The Netherlands}}%
}
\affil[4]{
    \orgname{Vrije Universiteit Amsterdam},
    \orgaddress{\street{De Boelelaan 1111}, \postcode{1081 HV}, \city{Amsterdam}, \country{The Netherlands}}%
}

\abstract{
    Supervisory controllers control cyber-physical systems to ensure their correct and safe operation.
    Synthesis-based engineering (SBE) is an approach to largely automate their design and implementation.
    SBE combines model-based engineering with computer-aided design, allowing engineers to focus on `what' the system should do (the requirements) rather than `how' it should do it (design and implementation).
    In the Eclipse Supervisory Control Engineering Toolkit (ESCET\texttrademark) open-source project, a community of users, researchers, and tool vendors jointly develop a toolkit to support the entire SBE process, particularly through the CIF modeling language and tools.
    In this paper, we first provide a description of CIF's symbolic supervisory controller synthesis algorithm, and thereby include aspects that are often omitted in the literature, but are of great practical relevance, such as the prevention of runtime errors, handling different types of requirements, and supporting input variables (to connect to external inputs).
    Secondly, we introduce and describe CIF's benchmark models, a collection of 23 freely available industrial and academic models of various sizes and complexities.
    Thirdly, we describe recent improvements between ESCET versions v0.8 (December 2022) and v4.0 (June 2024) that affect synthesis performance, evaluate them on our benchmark models, and show the current practical synthesis performance of CIF.
    Fourthly, we briefly look at multi-level synthesis, a non-monolithic synthesis approach, evaluate its gains, and show that while it can help to further improve synthesis performance, further performance improvements are still needed to synthesize complex models.
}


\maketitle

% configure layout
\flushbottom
\sloppy

\section{Introduction}

A supervisory controller, supervisor for short,
coordinates the behavior of a cyber-physical system according to discrete-event observations of the system's behavior.
Based on such observations, the supervisor decides which events the system can safely perform and which events must be disabled, because they would lead to violations of requirements or to a blocking state.
Engineering of supervisors is a challenging task, due to the high complexity of real-life discrete-event systems~\cite{Sanden2015,Fokkink2022,Fokkink2023mikroniek,Baubekova2024}.

Synthesis-based engineering (SBE) is an engineering approach to design and implement supervisors that combines model-based engineering with computer-aided design, to produce correct-by-construction controllers, by automating the engineering process as much as possible.
A main ingredient of SBE is supervisory controller synthesis~\cite{Ramadge1987}, a theory and technique for automatically deriving a model of a supervisor from discrete-event models of the uncontrolled system behavior and the system's requirements, such as functional or safety-related requirements.
This allows engineers to focus on \emph{what} the system should do (the requirements) rather than \emph{how} the controller should do it (the design and implementation).

The Eclipse Supervisory Control Engineering Toolkit (ESCET\texttrademark, pronounced \emph{\`es\`et}) project\footnote{See \url{https://eclipse.dev/escet}. `Eclipse', `Eclipse ESCET' and `ESCET' are trademarks of Eclipse Foundation, Inc.} \cite{Fokkink2023escet}, provides a model-based approach and industrially-applicable toolkit for the development of supervisors.
It targets the entire model-based engineering process for the development of supervisors, including modeling, synthesis, simulation-based validation and visualization, formal verification, real-time testing, and code generation.
The ESCET toolkit contains multiple modeling languages and associated tools.
In this paper, we focus on one of them, CIF\footnote{See \url{https://eclipse.dev/escet/cif}. `CIF' originally stood for `Compositional Interchange Format for hybrid systems', a formalism to interface multiple different discrete event, timed and hybrid formalisms, making it easier to transform them to each other. Although the CIF tools still feature various transformations to other formalisms, the role of CIF as an interface format has declined. While the name `CIF' stuck, it is no longer considered an acronym.} \cite{Beek2014}, which features an automata-based modeling language for convenient specification of large-scale systems and tools that support the entire SBE process.
The ESCET project also comprises Chi\footnote{See \url{https://eclipse.dev/escet/chi}.} \cite{Schiffelers2003}, a hybrid modeling language and toolset to analyze the performance of discrete event systems through simulation, which is beyond the scope of this paper.
And finally, the ESCET project features the ToolDef\footnote{See \url{https://eclipse.dev/escet/tooldef}.} scripting language for the definition and execution of model-based toolchains, useful for automating the execution of ESCET tools and for combining them.
We use ToolDef for the experiments in this paper.

The ESCET project, an Eclipse Foundation open-source project since 2020, builds upon decades of research and tool development at the Eindhoven University of Technology.
Vital for the evolution from an academic toolkit into an industrially-applicable toolkit are various years-long research collaborations with industry and government, such as with Rijkswaterstaat~\cite{Reijnen2017,Reijnen2018a,Fokkink2022}, ASML \cite{Sanden2015,Vos2020,Thuijsman2021} and Philips \cite{Theunissen2014,Theunissen2015}.
Rijkswaterstaat, part of the Dutch Ministry of Infrastructure and Water Management, is responsible for infrastructure in the Netherlands, including roads, bridges, tunnels, and waterway locks.
ASML is a leading company in the semiconductor industry and provides chipmakers with all they need to mass produce patterns on silicon through lithography.
Philips is a leading company in healthcare technology and provides, among others, medical equipment such as MRI and X-ray machines.
The quality of supervisory control software for such systems impacts their availability, reliability, and safety.
Synthesis-based engineering allows for automation, modularization and standardization, increasing quality and evolvability, and decreasing life-cycle costs~\cite{Theunissen2014,Baubekova2024}.

Supported by the Eclipse Foundation's principles of transparency, openness, meritocracy, and vendor-neutrality, the ESCET project aims to be an open environment and a growing community.
It allows interested parties, such as academic and applied research institutes, industrial partners, and tool vendors, to collaborate on and profit from further tool development for the model-based construction of supervisors.
Furthermore, the project's open nature allows any vendor to develop commercial tool support.

The increasing complexity of industrial systems for which supervisory control synthesis can be applied, as well as the intrinsic complexity of this synthesis method, make it essential to push the performance of the ESCET toolkit where possible.
Moreover, to facilitate seamless integration into industrial design processes, usability is key.

The use of extended finite automata (EFAs) for modeling~\cite{Miremadi2008,Miremadi2011}, in combination with the use of a symbolic supervisory controller synthesis algorithm that uses Binary Decision Diagrams (BDDs) as data structure~\cite{Ouedraogo2011}, has been instrumental in efficiently modeling and synthesizing monolithic (a.k.a., single) supervisory controllers for industrial-size cases~\cite{Fei2014}.

This paper reports on some crucial recent developments of the ESCET toolkit that have considerably improved its capability to deal with the complexity of real-life systems.
It also shows how cross-fertilization with the model checking domain has played a significant role in boosting the performance of state-of-the-art supervisory controller synthesis tools.
More specifically, this paper has four main contributions:
\begin{enumerate}
    \itemsep 0em
    \item We provide a description of CIF's symbolic supervisory controller synthesis algorithm, and thereby include aspects that are often omitted in the literature, but are of great practical relevance, such as the prevention of runtime errors, handling different types of requirements, and supporting input variables (to connect to external inputs).
    This description assists researchers who wish to improve the algorithm.
    \item We introduce and describe CIF's set of benchmark models, a collection of 23 industrial and academic models of various sizes and complexities, that are freely available, allowing researchers to benchmark synthesis algorithms and their improvements.
    \item We describe recent improvements between ESCET versions v0.8 (December 2022) and v4.0 (June 2024) that affect synthesis performance, evaluate them on our set of benchmark models, and show the current practical synthesis performance of CIF.
    This helps researchers and practitioners to understand how much more scalable synthesis has become, as well as how scalable it currently is, from a practitioners point of view.
    \item We briefly look at multi-level synthesis, a non-monolithic synthesis approach, evaluate its gains, and show that while it can help to further improve synthesis performance, further performance improvements are still needed to synthesize complex models.
\end{enumerate}

The artifact accompanying this paper includes the models and scripts that can be used to reproduce our results~\cite{ArtifactOfThisPaper}.

\medskip

\textbf{Related work: }
Currently, the combination using EFAs for modeling with a BDD-based symbolic synthesis algorithm is employed by both Supremica~\cite{Malik2017} and CIF.
Other tools for synthesizing a single monolithic supervisory controller include DESTool \cite{Moor2008}, DESUMA \cite{Ricker2006}, and TCT \cite{Feng2006}.
These tools use finite automata instead of extended finite automata and do not employ a symbolic synthesis algorithm (using BDDs).
A comparison between these tools is provided in \cite{Reniers2018}.

An alternative approach to deal with the complexity of synthesis problems is to use non-monolithic synthesis approaches such as modular~\cite{Wonham1988,Queiroz2000,Malik2016}, decentralized~\cite{Lin1988,Rudie1991,Lee2022}, hierarchical~\cite{Zhong1990,Wong1996}, compositional~\cite{Flordal2007} and multi-level synthesis~\cite{Komenda2016}, and combinations of them~\cite{Schmidt2004,Hill2006}.
These approaches divide the synthesis problem, on different grounds, into a number of `smaller' synthesis problems, and thus result in multiple synthesized supervisors (one for each part of the system), which together ensure the correct and safe operation of the system.

The above-mentioned approaches all deal with the synthesis of maximally permissive, controllable, nonblocking, and safe supervisory controllers, where safety is specified by requirements (automata and propositional logic properties).
The synthesis of more involved properties, such as those described by temporal logic formulas, is receiving much attention in the literature (e.g., \cite{Hulst2017,Ehlers2014}).
So far, these approaches have not been applied in industrial-size cases.

For timed discrete event systems, a symbolic approach has been published recently for the synthesis of a supervisory controller for a plant described using timed automata \cite{Rashidinejad2024}.
The UPPAAL-Tiga tool allows for the solving of games based on timed game automata, resulting in a synthesized controller \cite{Behrmann2007}.
A recent overview of advances, including a systematic literature study, is provided by Fokkink et al.~\cite{Fokkink2024}.

\medskip

\textbf{Outline: }
We provide background information on supervisory controller synthesis and the SBE process in Section~\ref{sec:background}, before we introduce CIF's symbolic synthesis algorithm in Section~\ref{sec:cif-synth-algo}.
We then describe the recent improvements in Section~\ref{sec:improvements}, introduce the benchmarks in Section~\ref{sec:benchmarks}, and experimentally evaluate the improvements on these benchmarks in Section~\ref{sec:experiments}.
We further look at non-monolithic synthesis, and multi-level synthesis in particular, in Section~\ref{sec:non-monolithic}, before we conclude in Section~\ref{sec:conclusions}.

\section{Background}
\label{sec:background}

This section presents brief informal explanations of supervisory controller synthesis and synthesis-based engineering.
More precise definitions, especially with regard to extended finite automata and synthesis, will be provided in Section \ref{sec:cif-synth-algo}.

\subsection{Supervisory controller synthesis}

\begin{figure}[b!]
    \centering
    \begin{tikzpicture}
        \input{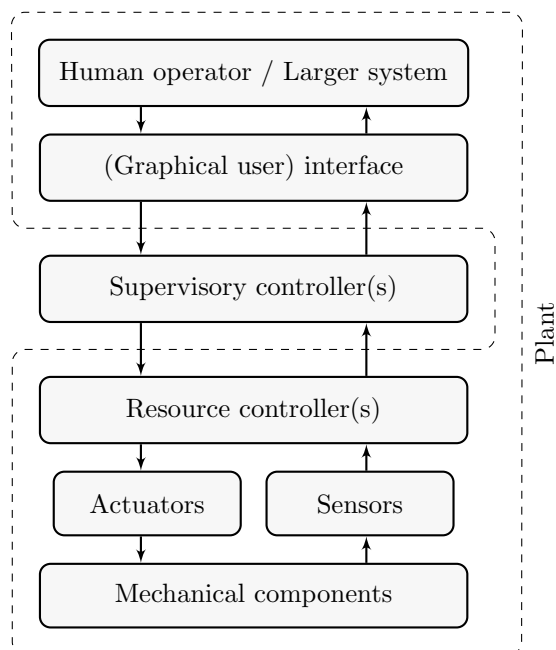}
    \end{tikzpicture}
    \caption{Structure of supervisory control. Part enclosed with dashed line is the `plant'. Adapted from a similar figure in~\cite{Fokkink2023escet}.}
    \label{fig:supervisory-control-structure}
\end{figure}

Figure~\ref{fig:supervisory-control-structure} depicts the general system structure for supervisory control.
A cyber-physical system consists of mechanical components to be controlled, such as robot arms, conveyor belts, and so on.
Actuators drive their operation, while sensors indicate their status.
Resource control provides low-level control, often offering more abstract actuator and sensor signals for higher levels of control to use.
Supervisors ensure actuator signals that would violate requirements are disabled.
Large systems may be divided into (layers of) subsystems, and supervisors can be present at each level, coordinating lower-level subsystems (only a single layer is depicted).
A (sub)system is often controlled by a human operator through a graphical user interface, or part of a larger system to which it is connected by an interface.
The part of the system that is controlled by the supervisor, so the environment of the supervisor, is called the \emph{plant} or \emph{uncontrolled system}.
In the figure it is the part of the system enclosed within the dashed line.
This may include higher layers, as for instance warning lights of the GUI may need to be controlled.
The uncontrolled system and supervisor are together called the \emph{controlled system}.

Supervisory controller synthesis~\cite{Ramadge1987,Wonham2018} automatically generates a correct-by-construction supervisor model for a discrete-event system, given precise descriptions of the behavior of the plant components and the (safety) requirements for the desired system behavior.
These can be specified conveniently as extended finite automata (EFAs), i.e., \emph{automata} with \emph{locations} that model the local states of a system component, and \emph{edges} between the locations that model how a component can transition from one such local state to another.
EFAs are decorated with \emph{variables} that allow us to model data.
For an example of an EFA, see Figure~\ref{fig:linearization-pre1} later in this paper.
Various parts of the model may then carry \emph{expressions} that express conditions on states of the system (i.e., over the variables and locations of the automata) or calculations over the system state.
Crucial concepts are \emph{guards} on edges (i.e., boolean conditions that must be \emph{true} for the corresponding edge to be enabled), \emph{updates} on edges (specifying how the system state evolves as the result of taking an edge), and \emph{invariants} (i.e., boolean conditions that must remain \emph{true} during system execution, thereby restricting the reachable or allowed states) \cite{Skoldstam2007,Markovski2010}.

Synthesis considers the parallel composition of the \emph{plant} automata that specify which behavior is physically possible (at the individual components of the plant, as well as the interactions between the plant components), together with the \emph{requirement} automata that specify which system behavior is allowed.
That is, these automata synchronize on shared events, meaning these events must be executed simultaneously by all participating automata.
Multiple automata may thus participate in a single \emph{transition} for an event, taking the system from its current state, the \emph{source} state, to the \emph{target} state of the transition.
Thereby, each participating automaton takes an enabled edge for that event from its current location, performs the updates of its edge, and goes to the target location of the edge.
If for any participating plant automaton there is no edge for the event in the current location of the automaton, none of the present edges for the event in that location are enabled, or the event is otherwise restricted by a plant invariant, then no transition is possible for the event in the entire system, and the event is considered to be physically impossible in the current system state.
If, on the other hand, an event is similarly not possible only due to participating requirement automata, or is restricted by a requirement invariant, it is physically possible but must be disabled by the synthesized supervisor to ensure \emph{safety}.

Controllable events (such as output signals to actuators) can be prevented by a supervisor, but uncontrollable events (such as input signals from sensors) cannot.
To ensure \emph{controllability}, if an uncontrollable event must be prevented, the supervisor makes the system state where it occurs unreachable by disabling all controllable events leading to it.
Moreover, if an uncontrollable event leads to such a state, then the origin state of this event must also be made unreachable.

If safety of, for instance, a drawbridge is ensured by forcing it to remain raised forever, it is useless for road traffic.
Therefore locations of the plant and requirement automata can be marked, for instance locations where the bridge deck is lowered, the barriers are open, and the signals are green.
A marked state in the parallel composition means that all individual components are in a marked location, in this case allowing traffic to proceed over the bridge.
The supervisor must guarantee that the plant can always reach a marked system state, by disabling (events leading to) states that violate this property.
Only then is the controlled system \emph{nonblocking}.

Supervisory controller synthesis ensures \emph{safety}, \emph{controllability} and \emph{nonblockingness} of a system with respect to its requirements, accounting for all possible behavior, also disabling events that lead to problems such as blocking behavior or requirement violations much later in the system's execution.
It does so by restricting as little behavior as possible, thereby ensuring \emph{maximal permissiveness}.

The use of EFAs also allows extracting and representing the synthesized supervisor compactly and intuitively~\cite{Miremadi2008,Miremadi2011}.
CIF represents supervisor models as the collection of the provided plant and requirement models together with the addition of a single supervisor EFA.
This supervisor EFA has for each controllable event an additional guard, which represents the extra restrictions that the supervisor imposes in order to ensure safety, controllability and nonblockingness.

\subsection{Synthesis-based engineering process}

The CIF language and tools, as part of the ESCET toolkit, together enable a synthesis-based engineering approach and support its entire process.
The CIF language is a declarative modeling language for the specification of discrete-event, timed, and hybrid systems as a collection of synchronizing automata.
It can be used to specify various different models during the entire development process, including plants and requirements models as well as simulation models, as it has a rich set of concepts.
This prevents the need to use multiple languages during the engineering process.

The CIF tools support the various steps of the synthesis-based development process, including among others specification, supervisory controller synthesis, simulation-based validation and visualization, verification, real-time testing, and code generation.
The toolkit has a strong focus on industrial application, with, e.g., modeling convenience, efficient algorithms, and checks for common mistakes, as well as extensive documentation, tutorials, and examples.
Some CIF tools only support a subset of the CIF language concepts.

\begin{figure}[t!]
    \begin{center}
        \begin{tikzpicture}
            \input{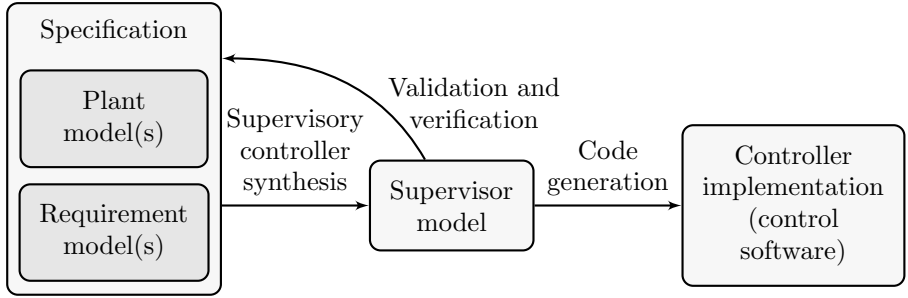}
        \end{tikzpicture}
    \end{center}
    \vspace{-3mm}
    \caption{Simplified representation of ESCET's synthesis-based engineering process.}
    \label{fig:toolchain}
\end{figure}

We briefly introduce CIF's synthesis-based engineering process and associated tools\footnote{See for more detailed information the CIF website: \url{https://eclipse.dev/escet/cif/synthesis-based-engineering/} and \url{https://eclipse.dev/escet/cif/tools}.}, as shown in simplified form in Figure~\ref{fig:toolchain}.
It starts with a model-based specification, consisting of plant and requirement models, modeled as EFAs with invariants.
To these models, supervisory controller synthesis is applied, resulting in a model of the supervisor.
The ESCET toolkit supports synthesis both with its own synthesis tools and by a transformation to Supremica, for a subset of CIF models.

Synthesis ensures that all specified requirements are satisfied by the synthesized supervisor, and thus the requirements are correctly enforced.
Validation is supported by CIF's automated or interactive simulation and visualization.
It helps to determine whether the specified plants and requirements, and by extension the supervisor, achieve the desired system behavior.
It allows us to check that the plants properly represent the possible behavior of the system as understood by the engineers, and that the specified requirements are the intended requirements and lead to the desired behavior.
Verification, such as model checking, supported through transformations to UPPAAL \cite{Behrmann2006,Agut2011} and mCRL2~\cite{Bunte2019,Reniers2024}, can also be used to spot flaws in the plant and requirement models.
It can moreover be employed to check other requirements not yet supported by synthesis, including stronger liveness guarantees, timing properties, and some additional properties that must hold for the supervisor to be properly implemented as a controller~\cite{Reijnen2019}.

An implementation of the controller can then be obtained automatically from a model of the supervisor, by generating code for its control software.
The CIF tools support code generation for multiple languages and platforms, including Java, JavaScript, C, and Simulink, as well as Programmable Logic Controller (PLC) code according to the IEC standard 61131-3 for multiple vendors.

\section{Symbolic supervisory controller synthesis}
\label{sec:cif-synth-algo}

The `data-based synthesis tool' is CIF's primary synthesis tool.
It is based on the symbolic supervisory controller synthesis algorithm of Ouedraogo et al.~\cite{Ouedraogo2011}.
It iteratively strengthens the guard predicates on edges so that forbidden states become unreachable in the controlled system.
The use of a symbolic algorithm prevents that all states of the system have to be explicitly represented, which would be infeasible for the kind of real-world systems we consider in Section~\ref{sec:benchmarks}~\cite{Burch1992}.
Symbolic synthesis therefore represents a major step forward for the industrial applicability of supervisory controller synthesis, by allowing efficient synthesis of plants and requirements modeled as EFAs~\cite{Fei2014}, as we will also show in Section~\ref{sec:experiments}.

In this section we provide a description of CIF's symbolic synthesis algorithm.
Most parts have been presented separately in the literature before.
We describe these parts in some detail and integrate them into a single description of the synthesis algorithm.
Furthermore, we also include aspects that are often omitted in the literature~\cite{Ouedraogo2011,Thuijsman2021}, but are of great practical relevance, such as the prevention of runtime errors, handling different types of requirements, and supporting input variables (to connect to external inputs).
This description can assist researchers to better understand the algorithm, so that they can improve upon it.
We leave out most optional parts, that can be configured through the synthesis tool's many options.

\begin{figure}[t!]
    \centering
    \begin{tikzpicture}
        \input{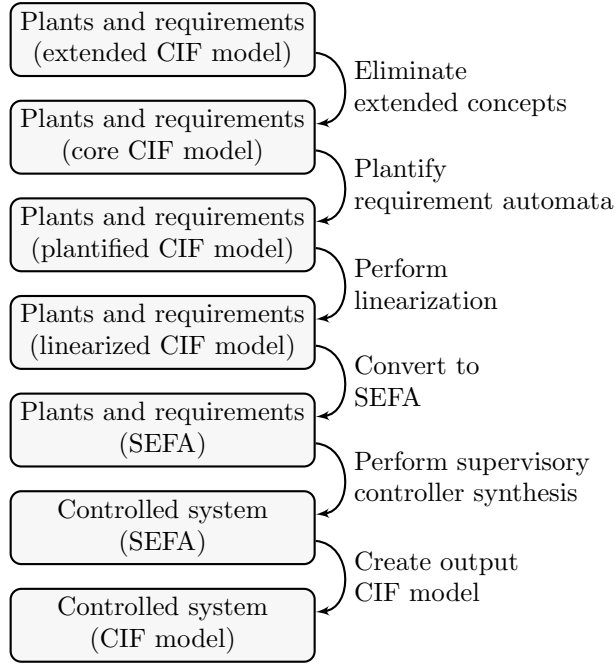}
    \end{tikzpicture}
    \caption{Overview of the steps involved in CIF's symbolic supervisory controller synthesis tool.}
    \label{fig:cif-synth-algo-overview}
\end{figure}

Figure~\ref{fig:cif-synth-algo-overview} shows an overview of the steps involved in CIF's symbolic synthesis tool.
A user makes a model with the plants and requirements, making use of CIF's various modeling concepts, and some of the extended concepts are eliminated to core concepts, to simplify subsequent steps, as described in Section~\ref{sec:cif-models}.
In the model with core concepts, to further reduce the number of concepts, the requirement automata -- but not the requirement invariants -- are `plantified' to plant automata and state/event exclusion requirements, as described in Section~\ref{sec:plantify}.
The (partially)
plantified model is linearized to eliminate synchronization between automata, which results in a linearized CIF model, with an automaton with a single location and self-loop edges, as described in Section~\ref{sec:linearization}.
The linearized model is then converted to a Symbolic EFA (SEFA).
We explain the symbolic encoding of predicates using Binary Decision Diagrams (BDDs) in Section~\ref{sec:bdds}, introduce SEFAs in Section~\ref{sec:sefa}, discuss the computation of SEFA transitions
in Section~\ref{sec:edge-apply}, and detail the conversion from linearized models to SEFAs in Section~\ref{sec:linearized-model-to-sefa}.
The SEFA is used to perform synthesis, resulting in a SEFA of the controlled (or supervised) system, as described in Section~\ref{sec:symbolic-synth-algo}.
Finally, the output CIF model is constructed from the controlled-system SEFA, as described in Section~\ref{sec:cif-model-creation}.

\subsection{CIF models}
\label{sec:cif-models}

We start at the beginning, with the plants and requirements model that an engineer makes, and that forms the input to the synthesis process.
We discuss what concepts are used in CIF to model these plants and requirements, omitting concepts that are not supported by synthesis.
Formal definitions, with regard to symbolic EFAs, will be presented in Section \ref{sec:sefa}.

\textbf{Automata:} CIF allows modeling the system as extended finite automata (EFAs).
An \emph{automaton} consists of one or more \emph{locations}, and is always exactly in one location at any time, its \emph{current} location.
An automaton further has directed \emph{edges} between the locations.
Each edge has a \emph{source} location and a \emph{target} location.
If an edge has the same source and target location, it is called a \emph{self-loop} edge.
The edges are labeled with \emph{events}.
A single edge labeled with multiple events is semantically equivalent to multiple edges between the same source and target location, each with one of the events.

\textbf{Discrete variables:} Automata may optionally declare \emph{discrete variables}, often simply called `variables'.
Each variable has an associated finite data type.
CIF has many data types, but for synthesis only booleans, ranged integers (non-negative numbers in an explicitly specified finite range) and enumerations are supported.

\textbf{States:} The \emph{state} of a CIF automaton consists of its current location and values for each of its variables.
A (global system) state of a CIF model consists of the combined (local) states of each of its automata.

\textbf{Guards:} An edge may have a \emph{guard} that indicates a restriction on when the edge may be taken.
If no guard is given, it is implicitly \emph{true}.
A guard may refer to any variable in the model, so not only of the automaton in which it is specified.
A guard is evaluated over the system state before the corresponding edge is taken.

\textbf{Updates:} An edge may also have any number of \emph{updates} in the form of assignments that give new values to variables.
For instance, an \emph{assignment} $x := 5$ assigns the value $5$ to $x$, and $x := x + 1$ increments the value of $x$ by one.
CIF has the `global read, local write' principle, meaning that all automata may read the values of all variables of the system, in for instance guards and right-hand sides of updates, but only the automaton in which the variable is declared may assign it new values.

\textbf{Initialization:} CIF allows any number of locations of an automaton to be designated as \emph{initial} locations.
A location may also be conditionally initial, by indicating an initialization predicate for the location, for instance depending on the initial values of variables, or on the initial locations of other automata.
Each variable is given one or more potential initial values.
The initial locations and initial values of variables can be further restricted by initialization predicates specified in components (inside and outside automata).
A system state is initial if each automaton is in an initial location, each variable has one of its possible initial values, and the state satisfies all initialization predicates.

\textbf{Marking:} CIF similarly allows any number of locations of an automaton to be \emph{marked}, and they may also be conditionally marked.
Marker predicates specified in components may further restrict the marking of both locations and variables.
A system state is marked if each automaton is in a marked location and the state satisfies all marker predicates.

\textbf{Transitions:} \emph{Transitions} between system states are possible by taking edges of automata.
If multiple automata share an event, the automata \emph{synchronize} on that event.
That is, they must all take an edge for that event, at the same time, to together perform a transition to a new system state.
In other words, each synchronizing automaton must have an edge for that event in its current location, and the guard of this edge must hold in the current system state.
Then each automaton takes one edge for the event, and they all go to the target location of their own edge.
Furthermore, the assignments of all edges are performed to determine the new state that is reached.
The right-hand sides of the assignments are all evaluated in the source state of the transition, making it irrelevant in which order the assignments are performed.

\textbf{Alphabets:} The \emph{alphabet} of an automaton defines the events over which the automaton synchronizes.
It can be explicitly specified, and otherwise it is implicitly equal to all the events that are used on the edges of the automaton.

\textbf{Runtime errors:} For example, given a variable with integer \emph{range} $[0,5]$, an assignment $x := x + 1$ for variable $x$ may lead to a \emph{runtime error}.
If $x = 5$, then after the assignment $x = 6$, which is outside the variable's range, causing a runtime error.
Synthesis in CIF has the implicit requirement that all variables must always be within their range, and that assignments must never result in runtime errors.
Synthesis thus enforces that variables are kept within their ranges.

\textbf{Invariants:} CIF furthermore supports two types of \emph{invariants}: \emph{state invariants} and \emph{state/event exclusion invariants}.
A state invariant indicates a predicate that must always hold, in all system states.
State/event exclusion invariants indicate a predicate that must hold, for a certain event to be allowed or disallowed.
They thus restrict the system states from which the event can be taken (plant invariant) or is allowed to be taken (requirement invariant).
For instance, `$e~\texttt{needs}~x = 1$' is a state/event exclusion invariant that indicates that event $e$ may only occur in states where $x = 1$ holds.
And `$x \neq 1~\texttt{disables}~e$' indicates that event $e$ may not occur in states where $x \neq 1$ holds.
These two state/event exclusion invariants are equivalent, and the one can be rewritten into the other, and vice versa.

\textbf{Plants:} For synthesis, all automata and invariants are labeled as being part of the \emph{plant} or being a \emph{requirement}.
\emph{Plant EFAs} indicate the possible behavior of the uncontrolled system.
Plant invariants restrict this possible behavior: \emph{state plant invariants} indicate impossible system states, and \emph{state/event exclusion invariants} indicate that certain events are impossible in certain system states.
For instance, an elevator could have two limit sensors that indicate whether it is at the bottom or at the top.
A state plant invariant could indicate that it is impossible for both sensors to be on at the same time, as the elevator cannot be both at the bottom and the top.
Alternatively, a state/event exclusion invariant could indicate that it is impossible for the top sensor to go on when the elevator is at the bottom.

\textbf{Requirements:} Requirements indicate constraints on the possible system behavior.
\emph{Requirement EFAs} are especially useful to indicate allowed orders of events, such as the way products must flow through a system.
\emph{State requirement invariants} indicate system states that may never be reached, such as two robots that are at the same position and would collide, causing damage.
\emph{State/event exclusion requirement invariants} indicate that certain events may not be performed in certain system states, such as opening a bridge when the traffic light is green and road traffic may thus still be driving over the bridge.

\textbf{Input variables:} CIF also features \emph{input variables}, which represent external values.
For synthesis, they must also have a finite data type, and their supported types are the same as for discrete variables.
Input variables can be read globally, in guards, right-hand sides of assignments, initialization predicates, marker predicates, invariants, and so on, like for discrete variables.
However, input variables cannot be assigned a value.
Instead, their value is determined by the environment of the model, and could change at any time.
They could, for instance, be connected to input ports of the system, such as input ports of a PLC.

\textbf{Extensions:} CIF offers various other language features that make it easier to model, such as \emph{constants} (variables with a fixed value), \emph{algebraic variables} (variables declaratively defined as an expression over the system's state), \emph{groups} to group together automata, \emph{imports} to allow reusing partial specifications, parameterized \emph{component and group definitions} that can be instantiated any number of times, and so on.
These concepts can be seen as syntactic sugar, extended concepts that can be eliminated into core concepts: automaton definitions can be instantiated to automata, constants and algebraic variables can be inlined, and so on.
Eliminating such extended concepts preserves the semantics of the model, while possibly easing subsequent steps.

\subsection{Plantification of requirement automata}
\label{sec:plantify}

Requirement automata are not taken into account directly during linearization, but are `plantified' to reduce the number of concepts to consider during synthesis.
They are converted to plant automata and state/event requirement invariants.
Note that in this step we only partially plantify the requirements, namely only the requirement automata.
The requirement invariants are not plantified (we will come back to this later).

More concretely, each requirement automaton is relabeled as a plant automaton that observes the system state, without restricting it.
To ensure that the automaton imposes no restrictions, for each event in the requirement automaton's alphabet, for each location without edges for that event, a self-loop edge is added to ensure the event is not restricted in that location.
If a location already has one or more edges for the event, a self-loop edge is added with the negation of the union of the guards of the existing edges for the event in that location.
This way, the plantified requirement automaton does not restrict the behavior of the system, and can be considered as a plant automaton.

To ensure that the restrictions of the requirement automaton are still taken into account during synthesis, for each added edge a state/event exclusion invariant is added that imposes the same restriction as its guard.

For instance, consider an event $e$ and two edges for that event in a location $l$ of a requirement automaton, with guards $g_1$ and $g_2$.
A self-loop edge is added to $l$ with event $e$ and guard $\lnot\, (g_1 \lor g_2)$.
Furthermore, a state/event exclusion requirement `$l\,\land\, \lnot\, (g_1 \lor g_2)~\texttt{disables}~e$' is added to the model.

\subsection{Linearization}
\label{sec:linearization}

The plantified CIF model may have many EFAs, which synchronize on shared events.
Based on work of Nadales Agut et al.~\cite{Nadales2011}, CIF's symbolic synthesis algorithm linearizes the CIF model, eliminating the parallel composition between the different EFAs, resulting in a single automaton with one location and only self-loop edges.

As part of this transformation, each EFA gets a location pointer variable, being a discrete variable that indicates its current location. 
Each edge is additionally guarded with a predicate that indicates whether the location pointer variable is equal to the value for its source location.
Each non-self-loop edge also gets an additional update that assigns the value matching the target location to the location pointer variable.

Parallel composition is then eliminated, by for each event combining the possible edges for that event from each synchronizing EFA, in all possible combinations.
For instance, if in one EFA there are three possible edges for the event, and in another there are four, then combined they get $3*4=12$ linearized edges.

The resulting single EFA thus has a single location and only self-loop edges.
The locations are then no longer relevant, as the state of the system is captured solely by the variables.
Internally, symbolic synthesis represents the model entirely as predicates over variables.

\begin{figure*}[t!]
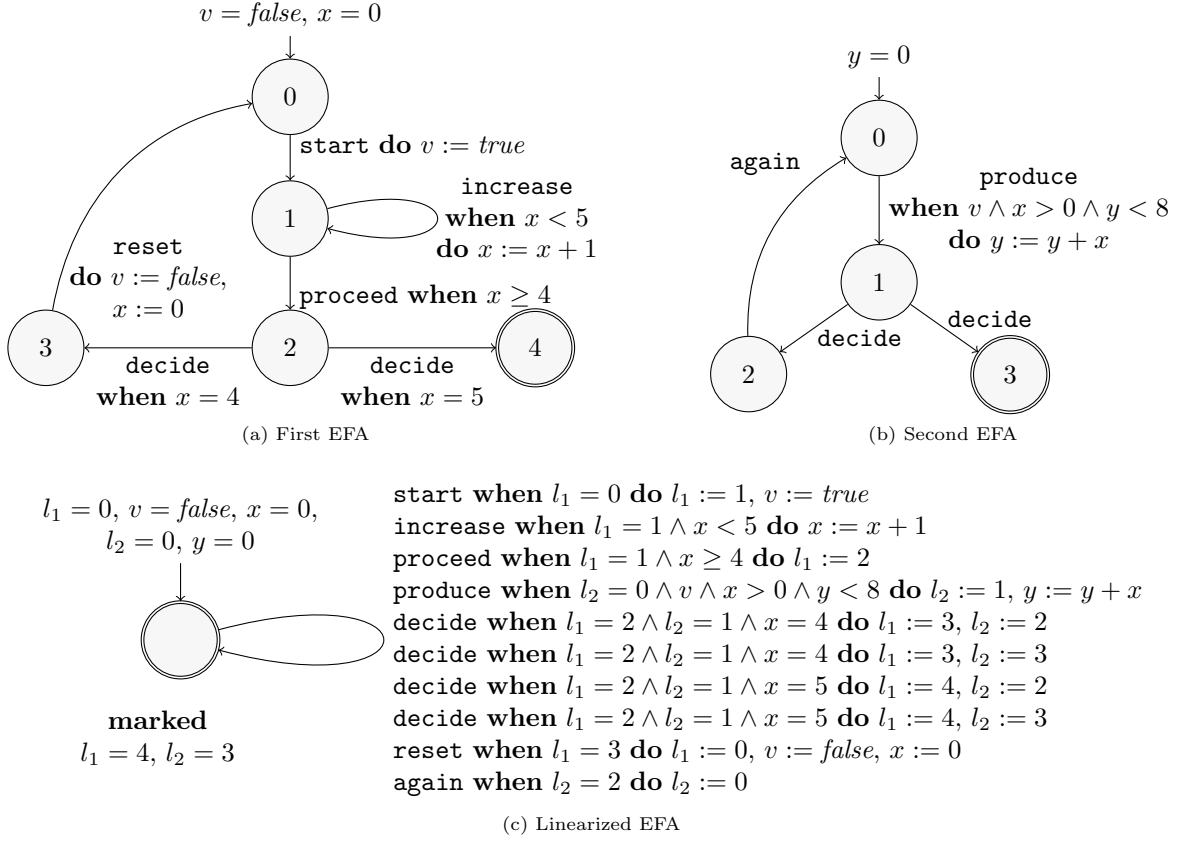

    \centering
    \subfloat[First EFA]{
        \begin{tikzpicture}[scale=0.5]
            \tikzset{
    auto,
    ->,
    node distance=3cm,
    align=center,
    uncontrollable/.style={densely dashed},
    font=\normalsize\normalfont,
    every state/.style={fill=black!3}
}

\node[state,initial above, initial distance=0.6cm, initial text={$v=\textit{false}$, $x=0$}, minimum size=1cm] (l0) {$0$};
\node[state,below=0.6cm of l0, minimum size=1cm] (l1) {$1$};
\node[state,below=0.7cm of l1, minimum size=1cm] (l2) {$2$};
\node[state,left=2.2cm of l2, minimum size=1cm] (l3) {$3$};
\node[state,right=2.2cm of l2, accepting, minimum size=1cm] (l4) {$4$};

\path[->] (l0) edge[] node[pos=0.2] {\texttt{start} \textbf{do} $v:=\textit{true}$} (l1);
\path[->] (l1) edge[loop right,min distance=4cm] node[xshift=-4pt] {\texttt{increase}\\\textbf{when} $x<5$\\\textbf{do} $x:=x+1$} (l1);
\path[->] (l1) edge[] node[pos=0.75] {\texttt{proceed} \textbf{when} $x\geq{}4$} (l2);
\path[->] (l2) edge[] node[below] {\texttt{decide}\\\textbf{when} $x=4$} (l3);
\path[->] (l2) edge[] node[below] {\texttt{decide}\\\textbf{when} $x=5$} (l4);
\path[->] (l3) edge[out=80,in=-170] node[right,pos=0.1] {\texttt{reset}\\\textbf{do} $v:=\textit{false}$,\\$x:=0$} (l0);
        \end{tikzpicture}
        \label{fig:linearization-pre1}
    }
    \hfil
    \subfloat[Second EFA]{
        \begin{tikzpicture}[scale=0.5]
            \input{images/linearization-pre2.tex}
        \end{tikzpicture}
        \label{fig:linearization-pre2}
    }
    \\
    \subfloat[Linearized EFA]{
        \begin{tikzpicture}[scale=0.5]
            \input{images/linearization-post.tex}
        \end{tikzpicture}
        \label{fig:linearization-post}
    }
    \caption{Example of two EFAs and their linearized form.}
    \label{fig:linearization}
\end{figure*}

Figure~\ref{fig:linearization} shows two EFAs (\ref{fig:linearization-pre1} and \ref{fig:linearization-pre2}) and their corresponding linearized EFA (\ref{fig:linearization-post}).
Locations are indicated by circles.
Marked locations have a double border.
Initial locations have an unconnected incoming arrow, with before it the initialization predicates and possible initial values of the variables.
A conditionally marked location is indicated by a predicate after \textbf{marked}, next to the location.
Marker predicates of components (outside of locations) are shown similarly.
Edges are indicated by arrows from circle to circle.
Edges are labeled with the event, and optionally a guard after \textbf{when} and updates after \textbf{do}.

In the linearized EFA several predicates are combined.
The initialization predicate of each location is made conditional on its corresponding value of the location pointer variable, and they are combined for the various locations of the automaton through disjunction (the example does not feature such conditional initial locations).
Similarly, marker predicates for the locations are combined per automaton.
And invariants, which may in CIF also be written inside locations, are also made conditional on being in the locations.

After linearization, there is a single EFA, with various linearized self-loop edges, initialization predicates, marker predicates, and invariants (the example has none).

\subsection{Binary Decision Diagrams}
\label{sec:bdds}

CIF's symbolic synthesis algorithm works on a symbolic representation of the linearized model, represented entirely as predicates over variables, which will be introduced in the next section.
But first, we introduce the Binary Decision Diagram (BDD) data structure, which CIF uses to symbolically represent the predicates representing (parts of) state spaces, and to perform computations on such predicates during synthesis (see Section~\ref{sec:symbolic-synth-algo}).
BDDs not only are compact in terms of storage, but also allow efficient manipulation~\cite{McMillan1993}.

BDDs are rooted, directed, acyclic graphs, with decision and terminal nodes.
There are two terminal nodes, one representing \emph{true} and one representing \emph{false}.
Each decision node is labeled with a boolean variable $v$, and has two outgoing edges to other nodes, one edge for when the variable is \emph{true} and one edge for when the variable is \emph{false}.
To evaluate a BDD for specific values of the variables, start at the root node, evaluate its variable, take the relevant edge, evaluate the variable of the next node that is reached, take the relevant edge, and so on, until a terminal node is reached, which indicates the resulting truth value of the predicate for the given values of the variables.

In our work, we use \emph{reduced ordered} BDDs (ROBDDs), which we will simply call BDDs from now on.
The variables are totally ordered using an ordering relation $<$, such that for any two variables $v_1 < v_2$, variable $v_1$ is placed closer to the root than $v_2$ in BDDs.
And the BDDs are reduced, meaning that they are always represented using a minimal number of nodes (with respect to a variable order), reusing existing nodes (and their subtrees toward the terminal nodes) and eliminating nodes of which both outgoing edges point to the same node.
By applying these restrictions whenever possible, a BDD always has a unique minimal representation, its canonical representation.

\begin{figure}[b!]
    \captionsetup[subfigure]{justification=centering}
    \centering
    \subfloat[Variable order\\$a\!<\!b\!<\!c\!<\!d$]{
        \begin{tikzpicture}
            \input{images/bdd-order1.tex}
        \end{tikzpicture}
        \label{fig:bdd-orders-good}
    }
    \quad\quad\quad\quad\quad
    \subfloat[Variable order\\$a\!<\!c\!<\!b\!<\!d$]{
        \begin{tikzpicture}
            \input{images/bdd-order2.tex}
        \end{tikzpicture}
        \label{fig:bdd-orders-bad}
    }
    \caption{Two BDDs representing $(a \land b) \lor (c \land d)$ for different variable orders.}
    \label{fig:bdd-orders}
\end{figure}

Figure~\ref{fig:bdd-orders} shows two BDDs for the same predicate over the same variables, but with two different variable orders.
Decision nodes are depicted as circles, labeled with boolean variables.
Terminal nodes are depicted as squares, labeled with F for \emph{false} and T for \emph{true}.
The root node is shown at the top, edges go down and are shown as arrows.
\emph{False} edges are depicted as dotted lines, \emph{true} edges as solid lines.

The size of a BDD is its number of decision nodes.
In this case, the same predicate can thus be represented with BDDs of sizes 4 (left) and 6 (right).
The variable order thus has an influence on the size of BDDs.
Since operations on BDDs are defined recursively over their structure, their computational effort is predominantly determined by the sizes of the BDDs on which they operate.
The variable order thus has a significant impact on both the amount of memory used to represent BDDs, and the time it takes to manipulate them.

CIF uses the JavaBDD library\footnote{See \url{https://github.com/com-github-javabdd/com.github.javabdd}.} for working with BDDs.
For synthesis, CIF supports variables with boolean, ranged integer, and enumeration data types.
To work with BDDs, each variable is represented using one or more boolean variables, its BDD variables.
For instance, a variable $x$ with the integer type with range $[0,5]$ can have six different values, which requires three boolean variables ($x_2$, $x_1$ and $x_0$), since $\lceil{}log_2(6)\rceil=3$.
Three variables can represent $2^3=8$ different values, of which in this case only six are used.
If $x_2=\textit{false}$, $x_1=\textit{true}$ and $x_0=\textit{true}$, $x$ has value $4\!*\!0 ~+~ 2\!*\!1 ~+~ 1\!*\!1 = 3$.
For each CIF variable $v$, also a variable $v^+$ is created, that allows representing the new value of the variable in update predicates. (This will be explained in more detail in the next section.)
The current and new value boolean variables are kept together in the variable order, and they are interleaved.
For instance, for CIF variable $x$, the boolean variables are in the following order: $x_0 < x_0^+ < x_1 < x_1^+ < x_2 < x_2^+$.

\subsection{Symbolic EFAs}
\label{sec:sefa}

As mentioned, CIF's symbolic synthesis algorithm works on a symbolic representation of the linearized model, represented entirely as predicates over variables.
We therefore introduce the concept of a \emph{symbolic EFA}, or \emph{SEFA}.
A SEFA is a 6-tuple $(V\!, D, \Sigma, E, p_0, p_m)$, with $V$ its finite set
of \emph{variables}, $D$ for each variable its finite domain of possible values,
$\Sigma$ its finite set of events called the \emph{alphabet}, $E$ its finite set of \emph{edges} (the \emph{transition relations}), $p_0$ a predicate denoting the \emph{initial} states, and $p_m$ a predicate denoting the \emph{marked} states.
We may sometimes refer to a SEFA as an EFA, if the difference is not relevant.

\textbf{Events:} Alphabet $\Sigma$ is partitioned into two disjoint sets: $\Sigma_c$ contains the controllable events and $\Sigma_u$ the uncontrollable ones.

\textbf{Variables:} $V$ contains $n\geq 0$ variables.
Each variable $v_i$, $1 \leq i \leq n$, has a value from its finite \emph{domain} of possible data values $D_i$.
$D$ is a function that for each variable $v_i$ gives it domain $D_i$, i.e., $D(v_i)=D_i$.
The set of possible states $X = D_1 \times D_2 \times \ldots \times D_n$ then gives all possible combinations of values for all of the variables, while a single state $x \in X$ is a tuple, a \emph{valuation} that associates with each variable $v_i$ a single value from $D_i$.

\textbf{Predicates:} A \emph{predicate} $p$ is a boolean expression over variables $V$\!, such as $a \land \lnot b$, $(c + 5) > (d ~mod~ 4)$, and so on.
Given a valuation $x \in X$, the variables $v \in V$ in a predicate $p$ can be replaced by their values in $x$, which we write as $p(x)$.
The resulting predicate is a boolean expression without variables, and thus its single truth value can be computed.
If it is \emph{true}, state $x$ is part of the set of states represented by $p$, while if it is \emph{false}, it is not part of the set of states.
A predicate $p$ thus symbolically represents a set of states $\{x \in X \,|\, p(x)\}$.
$p_0$ is the predicate that indicates the states in which the system may initially start, and $p_m$ is the predicate that indicates which states are marked.

\textbf{Edges:} Each edge $(g, r, \sigma, u) \in E$ consists of a guard predicate $g$, an error predicate $r$ (explained below), an event $\sigma \in \Sigma$, and an update $u$.
Sets $E_c, E_u \subseteq E$ represent a partitioning of $E$ based on the events of the edges, such that $E_c$ contains all edges for controllable events in $\Sigma_c$, and $E_u$ contains all edges for uncontrollable events in $\Sigma_u$.
Guard predicate $g$ indicates for which states the edge is enabled, and thus a transition is possible (typically to another state).

\textbf{Updates:} We introduce for each variable $v \in V$ its counterpart $v^+ \in V^+$, with $V^+ = \{v^+ \,|\, v \in V\}$, representing the new values of variables after a transition.
We furthermore define $V^* = V \cup V^+$, the set of all old and new state variables.
Update predicate $u$ over variables $V^*$ then indicates how the variables can be updated by taking the edge, as it indicates which old and new value combinations are possible for source and target states of transitions for the edge.
For instance, an assignment $a := a + 1$, which increments $a$ by one, can be represented as update predicate $a^+ \!= a + 1$, with $a^+ \in V^+$ and $a \in V$.
It allows all old and new value combinations where the new value of variable $a$ is one higher than its old value.

\textbf{Partial transition relations:} In the literature on supervisor synthesis, it is not uncommon to add for each variable $x$ on an edge that is not assigned a value, an update predicate $x^+ \!= x$ to $u$, to ensure that the variable remains unchanged~\cite{Vahidi2005,Thuijsman2021}.
We use partial transition relations, and thus do not need these extra restrictions, which keeps the update relations smaller.
We will come back to this in Section~\ref{sec:edge-apply}.

\textbf{Error predicates:} Error predicate $r$ indicates for which states the update $u$ goes outside the BDD-representable range of the variable.
For an integer variable $x$ with a BDD-representable range $[x_{min},x_{max}]$ and assignment $x := e$, with $e$ some expression, $r$ is $e < x_{min} \lor e > x_{max}$.
Note that for technical reasons, $r$ only keeps variables within their BDD-representable ranges.
It does not yet ensure that variables remain within their original CIF variable ranges; we will come back to that later in Section~\ref{sec:linearized-model-to-sefa}.

For instance, consider a variable $y$ with CIF variable range $[0,5]$, and update $y^+ \!= y + 1$.
Variable $y$ is represented using 3 BDD variables, with BDD-representable range $[0,7]$.
Error predicate $r$ is then $y+1 < 0 \lor y + 1 > 7$, since after the update $y+1$ is the new value of $y$ and if that is smaller than $0$ or larger than $7$, then it cannot be represented using the 3 BDD variables.
In this particular case, $y$ has BDD-representable range $[0,7]$, so $y + 1$ is in range $[1,8]$, which means $y+1 < 0$ is \emph{false}, and thus only $y + 1 > 7$ is relevant.
Error predicate $r$ thus keeps $y$ within BDD-representable range $[0,7]$, but not yet within CIF range $[0,5]$.

\textbf{Auxiliary definitions:} We also introduce function $\textit{vars}(p)$ that, given a predicate $p$ over $V$ or $V^*$, returns the variables in $V$ that occur in $p$.
Similarly, $\textit{vars}^+(p)$ over $V^+$ or $V^*$ gives the variables in $V^+$ that occur in $p$.
Finally, given a predicate $p$ over variables from $V$\!, a predicate $p[V:=V^+]$ over variables from $V^+$ is obtained by replacing each variable $v \in V$ by its counterpart $v^+ \in V^+$.

\smallskip

As an example, the SEFA for the linearized EFA of Figure~\ref{fig:linearization-post}:
{
    \begin{align*}
        V =&~~ \{l_1, v, x, l_2, y\}\\
        D =&~~ l_1\rightarrow\mathbb{N}_{0..4}, v\rightarrow\mathbb{B}, x\rightarrow\mathbb{N}_{0..5}, l_2\rightarrow\mathbb{N}_{0..3}, y\rightarrow\mathbb{N}_{0..12} \\
        \Sigma =&~~ \{\texttt{start}, \texttt{increase}, \texttt{proceed}, \texttt{decide}, \texttt{reset}, \texttt{produce}, \texttt{again}\} \\
        E =&~~ \{e_1, e_2, e_3, e_4, e_5, e_6, e_7, e_8, e_9, e_{10}\} \\
        e_1 =&~~ (l_1=0,~~ \textit{false},~~ \texttt{start},~~ l_1^+\!=1 \land v^+) \\
        e_2 =&~~ (l_1=1 \land x<5,~~ x+1>7,~~ \texttt{increase},~~ x^+\!=x+1)\\
        e_3 =&~~ (l_1=1 \land x\geq 4,~~ \textit{false},~~ \texttt{proceed},~~ l_1^+\!=2) \\
        e_4 =&~~ (l_2=0 \land v \land x>0 \land y<8,~~ y+x>15,~~ \texttt{produce},~~ l_2^+\!=1 \land y^+\!=y+x) \\
        e_5 =&~~ (l_1=2 \land l_2=1 \land x=4,~~ \textit{false},~~ \texttt{decide},~~ l_1^+\!=3 \land l_2^+\!=2) \\
        e_6 =&~~ (l_1=2 \land l_2=1 \land x=4,~~ \textit{false},~~ \texttt{decide},~~ l_1^+\!=3 \land l_2^+\!=3) \\
        e_7 =&~~ (l_1=2 \land l_2=1 \land x=5,~~ \textit{false},~~ \texttt{decide},~~ l_1^+\!=4 \land l_2^+\!=2) \\
        e_8 =&~~ (l_1=2 \land l_2=1 \land x=5,~~ \textit{false},~~ \texttt{decide},~~ l_1^+\!=4 \land l_2^+\!=3) \\
        e_9 =&~~ (l_1=3,~~ \textit{false},~~ \texttt{reset},~~ l_1^+\!=0 \land \lnot\, v^+ \land x^+\!=0) \\
        e_{10} =&~~ (l_2=2,~~ \textit{false},~~ \texttt{again},~~ l_2^+\!=0) \\
        p_0 =&~~ l_1=0 \land \lnot\, v \land x=0 \land l_2=0 \land y=0 \\
        p_m =&~~ l_1=4 \land l_2=3
    \end{align*}
}
$\mathbb{N}_{a..b}$ represents the set of natural numbers in the range $[a,b]$, and $\mathbb{B}$ represents the boolean values: $\{\textit{false}, \textit{true}\}$.
Variables $l_0$, $x$, $l_2$ and $y$ have domains $\mathbb{N}_{0..4}$, $\mathbb{N}_{0..5}$, $\mathbb{N}_{0..3}$ and $\mathbb{N}_{0..12}$, and BDD-representable ranges $[0,7]$, $[0,7]$, $[0,3]$ and $[0,15]$, respectively.
The upper bounds $7$ and $15$ come back in the error predicates.
Note that we simplified $v^+ \!= \textit{true}$ to $v^+$, $v^+ \!= \textit{false}$ to $\lnot\, v^+$ and $v \!= \textit{false}$ to $\lnot\, v$.
And we omitted parts of error predicates that are trivially \emph{false}.

\subsection{Computing SEFA transitions}
\label{sec:edge-apply}

From a set of source states, we may for a SEFA edge compute the possible transitions to target states, and vice versa.
Computing transitions is also called `applying edges', `taking edges', or `computing the (pre-)image', and is an essential ingredient in synthesis.
To apply edges efficiently, we use two BDD operations: \emph{relnext} for forward application and \emph{relprev} for backward application~\cite{Dijk2017}.

$\textit{relnext}(e, p)$ computes, given an edge $e=(g,r,\sigma,u)$ and a predicate $p$ over $V$\!, the predicate $p'$ over $V$ of states that can be reached from states in $p$ by taking edge $e$.
Predicate $p$ indicates the source states.
To take the edge, guard $g$ must hold, error predicate $r$ must not hold, and update $u$ then indicates the possible target states.
Thus, $p \land g \land \lnot\, r \land u$ indicates the possible combinations of source and target states according to edge $e$.
We precompute transition relation $t = g \land \lnot\, r \land u$ for each edge, to avoid repeatedly computing it.
\emph{relnext} then computes $(\exists_{\textit{vars}(t)}(p \land t))[V^+ := V]$.

\emph{relnext} thus first computes $p \land t$.
Then, to get only the possible target states as a predicate over $V^+$\!, we take the existential quantification~\cite{Burch1994} over $V$\!.
That is, predicate $p \,\land\, t$ is a predicate over $V^*$, and the existential quantification gives us a predicate over $V^+$, for any valuation of the variables in $V$ that satisfies $p \,\land\, t$.
We perform the quantification over a limited set of old state variables $vars(t)\subseteq V$\!, that contains only the old state variants of the variables that occur in $t$.
By not quantifying over all old state variables $V$\!, but only the relevant ones, allowing us to use a partial transition relation~\cite{Burch1991}, we achieve better performance.
It reduces the amount of quantification needed.
And it also avoids having to add $x^+ \!= x$ relations to $u$ that prevent unassigned variables from being unconstrained and thus getting any value.
This leads to smaller predicates and thus less work in computations.

After applying the edge, the target states should become the current states.
We therefore want to have a predicate over variables $V$\!, rather than over variables $V^+$.
Thus, \emph{relnext} performs substitution $[V^+ := V]$ to obtain predicate $p'$ over $V$\!, of states that can be reached from $p$ by taking edge $e$.

BDD operation $\textit{relprev}(e, p)$ is similar.
It computes for edge $e$ and predicate $p$ over $V$\!, the predicate $p'$ over $V$ of states from which by taking edge $e$ the states of $p$ can be reached.
That is, $\textit{relprev}(e, p)$ computes $\exists_{\textit{vars}^+(t)}(t \land p[V := V^+])$, with $\textit{vars}^+(t) \subseteq V^+$ the relevant new state variables.

By combining various operations into a single BDD operation, and by precomputing $t$, BDD operations \emph{relnext} and \emph{relprev} can efficiently apply SEFA edges.

\subsection{From linearized model to SEFA}
\label{sec:linearized-model-to-sefa}

The symbolic supervisory controller synthesis algorithm (described in Section \ref{sec:symbolic-synth-algo}) requires as input a single plant SEFA $P = (V\!, D, \Sigma, E, p_0, p_m)$, together with a predicate $p_f$ indicating states that are forbidden (e.g., by the requirements).
We now consider how this input can be derived from the linearized CIF model, with its discrete and input variables, linearized self-loop edges, initialization predicates, marker predicates, and invariants.

\textbf{Basic SEFA:} The discrete variables (including the added location pointer variables) and the input variables together determine $V$ and $D$.
The BDD variables corresponding to $V$ are ordered (see Section~\ref{sec:dcsh}).
The events used on the linearized edges determine alphabet $\Sigma$.
The linearized edges, with their events, guards, and updates are transformed to SEFA edges $E$.
To ensure that error predicates do not need to be considered explicitly in cases where we only use the guard and not the entire transition relation, we modify for each edge $e=(g,r,\sigma,u)$ its guard $g$ to $g \land \lnot\, r$.
The assignments of the linearized edges become updates $u$ of the edges, as described in Section~\ref{sec:sefa}.
The initialization predicates are combined through conjunction to form $p_0$.
The marker predicates are combined through conjunction to form $p_m$.

\textbf{Input variables:} Input variables can initially have any value, unless their initial values are constrained through initialization predicates.
They should be allowed to change to any other value at any time.
We therefore introduce, for each input variable $i$, a unique uncontrollable event $e_i$ and add it to $\Sigma$.
And we also add a SEFA edge for it, with guard \emph{true}, error \emph{false}, event $e_i$, and update $i^+ \neq i$, that allows $i$ to change to any other value than the one it currently has.

\textbf{Integer variable ranges:} An integer-typed discrete or input variable with a CIF range $[0,5]$ allows six different values, but as we saw in Section~\ref{sec:bdds}, it is represented using three BDD variables that can represent $2^3=8$ different values (BDD-representable range $[0,7]$).
To ensure that the variable never gets the two extra values, we forbid all states where it has a value outside of its CIF range.
That is, for every integer variable $x$ with CIF range $[x_{min},x_{max}]$, we add a state requirement invariant $x_{min} \leq x \land x \leq x_{max}$. 
Note that this is thus in addition to error predicates $r$ of edges (see Section~\ref{sec:sefa}), which prevent going outside of the BDD-representable range, but do not prevent using out-of-CIF-range values that fall within the BDD-representable ranges.
Furthermore, error predicates $r$ of edges do not ensure that the initial state satisfies integer variable ranges.

\textbf{Runtime errors for uncontrollable events:} Additionally, to ensure \emph{controllability}, for each edge with an uncontrollable event $\sigma \in \Sigma_u$, a state/event exclusion requirement invariant $\sigma ~\texttt{needs}~ g \land \lnot\, r$ is added.
The state/event exclusion invariant will ensure that $g \land r$ is added to $p_f$ (see below), thus ensuring that all states where uncontrollable event $\sigma$ is enabled and would lead to a runtime error are prevented from being reached.
Preventing runtime errors for controllable events will be discussed later, in Section~\ref{sec:symbolic-synth-algo}.

\textbf{State plant invariants:} The state plant invariants are incorporated in the initialization predicate and the guards of the edges.
We first compute $p_p$, the full restriction imposed by all state plant invariants, by combining these predicates with conjunction.
We change $p_0$ to $p_0 \land p_p$ so that the system can only start in states where $p_p$ holds.
We must then prevent going from a state where $p_p$ holds to a state where it does not hold.
If $p_p$ holds before taking an edge $e$, then it should still hold after taking the edge.
We compute from which states $s$, after taking edge $e$, we can reach states where $p_p$ still holds, by applying the edge backwards to $p_p$, i.e., $s=\textit{relprev}(e, p_p)$.
Finally, we update guard $g$ of $e$ to $g \land s$, to ensure we only end up in states where $p_p$ holds after taking the edge.
We thus start in states where $p_p$ holds, by adapting $p_0$, and stay in states where $p_p$ holds, by adapting the guards of all edges.
Note that we apply the optimization of Thuijsman et al.~\cite{Thuijsman2021}, who first check whether $s$ is already implied by $g \land p_p$, and forgo updating $g$ if that check holds.

\textbf{State requirement invariants:} The state requirement invariants indicate conditions that must always hold in every state.
Their inverse conditions indicate states that must never be reached.
We first compute the full restriction $p_r$ imposed by all state requirement invariants, by combining their predicates with conjunction.
We then initialize $p_f$, the forbidden states predicate, to $p_f = \lnot\, p_r$.

\textbf{State/event exclusion plant invariants:} The state/event exclusion plant invariants are incorporated into the edges.
Each \texttt{disables} state/event exclusion plant invariant is transformed to a corresponding \texttt{needs} invariant.
Each `$\sigma~\texttt{needs}~p$' state/event exclusion plant invariant is applied to each edge $(g,r,\sigma,u)$ by adapting $g$ to $g \land p$.

\textbf{State/event exclusion requirement invariants:} The state/event exclusion requirement invariants are mostly considered in the same way.
The \texttt{disables} invariants become \texttt{needs} invariants.
The guards of edges with controllable events are restricted in the same way as for state/event exclusion plant invariants.
Edges with uncontrollable events are handled differently, as synthesis may not restrict the guards of uncontrollable edges, to ensure \emph{controllability} of the resulting controlled system.
For each state/event exclusion requirement invariant `$\sigma ~\texttt{needs}~ p$', and each edge $(g,r,\sigma,u)$ with $\sigma\in\Sigma_u$, we compute the states $b=g \land \lnot\, p$, where guard $g$ holds, but the invariant does not hold ($\lnot\, p$).
Such states $b$ would allow the uncontrollable event, but do not satisfy the requirement, and since a supervisor cannot prevent uncontrollable event transitions, states where $b$ holds should never be reached.
Therefore, we update $p_f$ to $p_f \lor b$.

\subsection{Symbolic supervisory controller synthesis algorithm}
\label{sec:symbolic-synth-algo}

With these definitions, we define CIF's symbolic supervisory controller synthesis algorithm, see Algorithm~\ref{alg:synth}, which is based on the symbolic supervisory controller synthesis algorithm of Ouedraogo et al.~\cite{Ouedraogo2011}.
Given a plant model and forbidden states, the algorithm computes the controlled system behavior, those states of the system that are safe (adhere to all the implicit and explicit requirements), nonblocking (a marked state can always be reached), controllable (no uncontrollable events that are possible in the plant are blocked in safe nonblocking states), and optionally also reachable (contains only states reachable from an initial state), while remaining maximally permissive (not restricting more of the behavior than is necessary to ensure safety, nonblockingness, controllability and reachability).
It does so using several fixed-point reachability computations that are repeated until the controlled system states remain stable.
The guards of the edges, as well as the initialization and marker predicates of the model, are then strengthened to ensure that the model starts within the controlled system behavior and remains within it at all times.
Below, we discuss Algorithm \ref{alg:synth} in more detail.

\begin{algorithm}[b!]
    \caption{Symbolic Supervisory Controller Synthesis (\textit{SSCS})}\label{alg:synth}
    \begin{algorithmic}[1]
        \Require Plant SEFA $P = (V\!, D, \Sigma, E, p_0, p_m)$ and forbidden states $p_f$ over $V$.
        \Ensure Controlled-system SEFA $S$.
        \State $C \gets \lnot\, p_f$
        \Repeat
            \State $C' \gets C$
            \State $C \gets \textit{BRS}(p_m, E, C)$
            \State $B \gets \textit{BRS}(\lnot\, C, E_u, \textit{true})$
            \State $C \gets \lnot B$
            \State $C \gets \textit{FRS}(p_0, E, C)$ \Comment{Optional step.}
        \Until{$C = C'$}
        \ForAll{$e \in E$ with $\sigma \in \Sigma_c$}
            \State $g \gets g \land \textit{relprev}(e, C)$
        \EndFor
        \State $S \gets (V\!, D, \Sigma, E, p_0 \land C, p_m \land C)$
    \end{algorithmic}
\end{algorithm}

\textbf{Input: }
The algorithm takes as input a plant SEFA $P$, together with a set of states $p_f$ that are forbidden (e.g., by the requirements).

\textbf{Computing the controlled system states (lines 1--8): }
The algorithm starts (line 1) by initializing the controlled-system states predicate $C$ to be those states that are not forbidden, to ensure only safe states are part of the controlled-system behavior.

It then repeatedly computes non-blocking, controllable, and reachable states until a fixed point is reached (lines 2--8).
For each iteration, it first preserves the current controlled-system states predicate $C$ in $C'$ (line 3).
It then computes the non-blocking states from the marked states $p_m$ using a backward reachability search on all edges $E$, restricted to stay within the current controlled-system states $C$, using $\textit{BRS}(p_m, E, C)$ from Algorithm~\ref{alg:reach} (which is later explained in more detail); and it updates the controlled-system states $C$ to only those non-blocking states (line 4).

Subsequently, it performs a second backward reachability search on the inverse of the controlled-system (good) states $C$ (line 5).
By pushing these bad states backwards through edges labeled with uncontrollable events, more bad states are computed ($B$).
No additional restrictions are imposed on this reachability computation (\emph{true}).
This second reachability computation ensures that states with a transition for an uncontrollable event to a bad state are in turn labeled bad,
and thus ensures that the controlled system is controllable.
It also prevents runtime errors for edges with uncontrollable events, through $r$ of each edge, and through range restrictions of $p_f$ being in $C$.
The bad states are inverted and become the updated controlled-system states (line 6).

Optionally, an additional forward reachability search can be performed from the initial states $p_0$ (line 7).
Using a forward reachability search, the controlled-system states in $C$ are also all reachable.
By using forward reachability, the result of synthesis is generally more intuitive for engineers, as control conditions do not restrict unreachable states.
Forward reachability may also positively or negatively impact synthesis performance, as it may lead to smaller or larger predicates for $C$, which affects the effort required for subsequent reachability computations.
The forward reachability search is performed from the initial states, using all the edges, and is restricted to the current controlled-system states.

The loop with the various reachability searches is repeated until a fixed point is reached ($C = C'$) (line 8).
At this point, $C$ contains the controlled-system states, namely all safe, controllable, non-blocking (and, optionally, reachable) states.
Note that the implementation in ESCET has an optimization to prevent unnecessary fixed-point computations, and stops earlier if possible.
This is similar to what is described later in Section~\ref{sec:early-fixed-point-detection} for applying edges, but then for applying multiple fixed-point computations.
Furthermore, the implementation in ESCET checks after each reachability computation whether the combination of the uncontrolled system initialization predicate and the current controlled-system predicate still allows for initialization in the controlled system.
If not, then the algorithm can also stop earlier.

\textbf{Computing the controlled system model (lines 9--12): }
In the remainder of the algorithm, the controlled system, or supervised system, $S$ is determined (lines 9--12).

First, the guards of the edges with controllable events are restricted to ensure that the controlled system stays within the controlled-system states $C$ (lines 9--11).
Each edge gets a strengthened guard by combining the edge's existing guard with those conditions under which only states in the controlled-system behavior can be reached by considering the update of the edge (line 10).
This also ensures that runtime errors are prevented for edges with controllable events, similar to preventing it for uncontrollable events before.

Finally, the controlled-system SEFA is constructed, using the updated edges $E$, and by restricting the initial and marked states to controlled-system states $C$ (line 12).

\begin{algorithm}[t!]
    \caption{Backward Reachability Search (\textit{BRS})}\label{alg:reach}
    \begin{algorithmic}[1]
        \Require Start predicate $P$, edges $E$, and restriction predicate $R$.
        \Ensure The coreachable states $P'$, that is, those states that can reach states in $P$ via edges in $E$, restricted to states in $R$.
        \Repeat
            \State $P' \gets P$
            \ForAll{$e\in E$}
                \State $P \gets P \lor \textit{relprevIntersection}(e, P, R)$
            \EndFor
        \Until{$P = P'$}
    \end{algorithmic}
\end{algorithm}

\medskip

\textbf{Computing (co)reachable states (Algorithm \ref{alg:reach}): }
The algorithm performs backward reachability searches using the $\textit{BRS}$ algorithm, see Algorithm~\ref{alg:reach}, to compute coreachable states, i.e., states from which other states can be reached.
As input, the algorithm gets a start predicate $P$, edges $E$, and a restriction predicate $R$.
It performs a fixed-point computation in the loop (lines 1--4).
In each iteration, it preserves the current predicate $P$ in $P'$ (line 2).
It then updates $P$, adding all states from which states in $P$ can be reached by edges from $E$, restricting the result to $R$ (line 3).
BDD operation $\textit{relprevIntersection}(e, P, R)$ implements $\textit{relprev}(e, P) \land R$ as a single BDD operation, for better performance.
The edges are repeatedly applied until a fixed point is reached ($P = P'$), when $P$ contains the coreachable states (line 4).
The implementation in ESCET is actually a bit smarter in detecting a fixed point, see Section~\ref{sec:early-fixed-point-detection}.

The $\textit{FRS}$ algorithm performs a forward reachability search to compute reachable states (rather than coreachable states).
It is similar, but applies the edges forward using \textit{relnextIntersection}, rather than backward using \textit{relprevIntersection}.

\subsection{Supervised system CIF model creation}
\label{sec:cif-model-creation}

The last step that the synthesis tool performs is to create the output CIF model that represents the controlled system.
It does not directly convert the controlled-system SEFA to a single CIF EFA as output of the synthesis algorithm.
Instead, a supervisor EFA is derived from it, which is put in parallel to the original plant model, and restricts the plant behavior.

If the controlled-system SEFA has no initial state ($p_0 = \textit{false}$), then synthesis completes by indicating to the user that the result is an empty supervisor and no CIF model is produced.
Otherwise, the plantified CIF model (see Figure~\ref{fig:cif-synth-algo-overview}) is used as a starting point to construct the output CIF model.
All plantified requirement automata are labeled as supervisor automata, to make it clear that they track additional plant states for the supervisor.
Subsequently, all requirement invariants are removed, as they will be enforced by the supervisor.
And the initialization predicate $p_0$ of the controlled-system SEFA is added as an initialization predicate to the output model.

Finally, an extra supervisor automaton is added to complete the controlled-system CIF model.
Its alphabet is equal to the controllable events $\Sigma_c$ of the controlled-system SEFA.
It has a single location that is initial and marked.
For each (controllable) event in its alphabet, it has a self-loop edge without updates.
Each edge has a guard that indicates the condition under which the controllable event is enabled in the controlled system, i.e., the disjunction of the guards of the edges for that event in the controlled-system SEFA.

\medskip

The supervisor guards may become quite complex predicates.
To make it easier to understand the supervisor, the supervisor guard predicates are simplified, similar to the work of Miremadi et al.~\cite{Miremadi2008,Miremadi2011}.
That is, we simplify the guard predicates to only those conditions that must be additionally enforced on top of the plant behavior and restrictions already enforced by the requirements.
The extra restrictions are obtained by simplifying the controlled system guards under the assumption of the plant guards, plant invariants, requirements, and the controlled-system behavior.
Similarly, the initialization predicate of the controlled-system model is simplified to only those additional restrictions not already expressed by the uncontrolled-system initialization and the plant state invariants.

If synthesis did not need to restrict any further behavior, the resulting controlled-system CIF model has an additional controlled system initialization predicate \emph{true} and the supervisor automaton also has only \emph{true} guards.
If synthesis does impose additional restrictions, for instance to prevent runtime errors, or ensure non-blockingness or controllability, the resulting model will have a non-\emph{true} initialization predicate and/or supervisor guards.

Simplification is performed by default, but can be disabled.
If it is enabled, the supervisor no longer enforces the requirements.
The requirement invariants are then preserved, rather than removed, and are relabeled as supervisor invariants.

\section{Technical improvements}
\label{sec:improvements}

We describe the improvements that were made to CIF's symbolic synthesis tool between ESCET releases v0.8 (December 2022) and v4.0 (June 2024), but only as far as they affect synthesis performance.
We forgo describing improvements that were subsequently made obsolete by improvements in newer releases.
Concretely, we cover the improvements released as part of ESCET releases v0.9, v0.10, v1.0, v2.0, v3.0 and v4.0.
As the ESCET project makes quarterly releases, these six releases span one and a half years of ESCET tool development, although some improvements were in preparation already before the v0.8 release.

We opt for release v4.0, as it is the latest version of Eclipse ESCET at the time of writing this paper, and for release v0.8 as it is old enough such that we can investigate recent performance improvements, while not being so old that it is difficult to compare the performance practically (see Section \ref{sec:experiments}) for our set of benchmark models (see Section \ref{sec:benchmarks}).

The various improvements likely differ substantially in the impact that they have on the synthesis performance, as they are not all equally profound.
We do explain all of them, given that in Section \ref{sec:experiments} we will experimentally show the cumulative performance benefits that they bring.

Note that in this paper we consider practical performance for specific models, rather than theoretical complexity for any input model.
The theoretical worst-case complexity of the symbolic supervisory controller synthesis algorithm of Ouedraogo et al.~is $O(|L|^2|D|^2)$, with $|L|$ the size of the location pointer domains and $|D|$ the size of the data domains~\cite{Ouedraogo2011}.
This worst-case complexity does not change due to the improvements described in this paper.
Instead, the improvements help to improve the practical performance of synthesis, making it possible to efficiently synthesize supervisors for complex real-world systems, despite the high theoretical worst-case complexity bound.
The improvements for instance include various heuristics that have been shown to work well in practice, but do not offer guarantees, as well as improvements that help for some models, but not for all of them.

The following improvements are discussed in this section:
\begin{enumerate}
    \itemsep 0em
    \item DCSH variable ordering for BDDs (Section \ref{sec:dcsh})
    \item Linearized variable ordering relations for BDDs (Section \ref{sec:linearized-var-order-relations})
    \item Per-event edge granularity (Section \ref{sec:per-event-edge-granularity})
    \item Efficient edge application (Section \ref{sec:efficient-edge-application})
    \item Early fixed-point detection (Section \ref{sec:early-fixed-point-detection})
    \item Invariants: improved state plant invariants enforcement (Section \ref{sec:improved-state-plant-inv-enforce})
\end{enumerate}

\begin{figure}[b!]
    \centering
    \begin{tikzpicture}
        \input{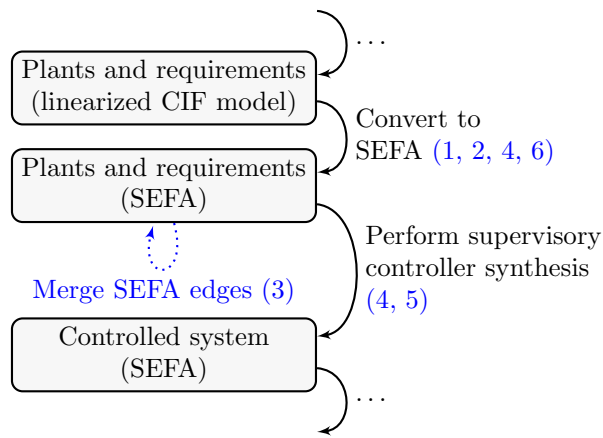}
    \end{tikzpicture}
    \caption{Synthesis performance improvements, positioned on the synthesis tool steps (only the relevant part is shown).}
    \label{fig:cif-synth-algo-improve}
\end{figure}

Figure~\ref{fig:cif-synth-algo-improve} positions these improvements, on the relevant part of the same overview presented earlier in Figure~\ref{fig:cif-synth-algo-overview}, as numbers in parentheses, and with an extra dotted/blue arrow.

\subsection{DCSH variable ordering for BDDs}
\label{sec:dcsh}

As mentioned in Section~\ref{sec:bdds}, the order of the boolean BDD variables is vital to the memory and running time characteristics of BDD representations and manipulations, as used by CIF's symbolic synthesis algorithm~\cite{Thuijsman2019}.
Heuristic variable ordering algorithms that exploit the inherent structure of the system modeled as EFAs are able to significantly reduce the synthesis effort~\cite{Lousberg2020}, especially for larger inputs, making synthesis applicable to more complex systems.

In CIF, the variable ordering algorithms order the CIF variables, and the order of the corresponding BDD variables is derived from that.
The BDD variables of a single CIF variable -- both the current and new state ones -- remain together, and are kept interleaved, as explained in Section~\ref{sec:bdds}.

The DSM-based Cuthill-McKee / Sloan variable ordering Heuristic (DCSH) is such a heuristic variable ordering algorithm, for which Lousberg et al.~experimentally showed a significant improvement on seven out of eleven of their case studies~\cite{Lousberg2020}.
The algorithm uses a Design Structure Matrix (DSM)~\cite{Eppinger2012}, a symmetric matrix in which relations between the variables of the linearized CIF model (see Figure~\ref{fig:cif-synth-algo-overview}) are encoded.
The rows and columns of the symmetric matrix are both labeled with the variables, and they are then reordered, to place closely related variables together, closer to the diagonal of the matrix.
DCSH uses two ordering heuristics, which operate on nodes of the undirected adjacency-graph equivalent of the matrix (each node represents a variable of the CIF model).
The two ordering algorithms that are used are the Sloan profile/wavefront-reducing node ordering heuristic~\cite{Sloan1989}, and a weighted version of the Cuthill-McKee bandwidth-reducing node ordering heuristic~\cite{Cuthill1969}.
For the orders produced by both algorithms, as well as the reverse of these two orders, DCSH then computes the Weighted Event Span (WES), a simple-to-compute metric that correlates to the computational effort of symbolic model checking~\cite{Meijer2016}, which involves similar computations as symbolic supervisor synthesis.
From the four variable orders (from the two algorithms, and those orders in reverse) the one with the best WES is selected as the result of the DCSH algorithm.

\begin{figure}[t!]
    \centering
    \subfloat[Before]{
        \begin{tikzpicture}
            \fill[black!0] (0mm,-0mm) rectangle (3mm,-3mm);
\fill[black!0] (3mm,-0mm) rectangle (6mm,-3mm);
\fill[black!0] (6mm,-0mm) rectangle (9mm,-3mm);
\fill[black!0] (9mm,-0mm) rectangle (12mm,-3mm);
\fill[black!0] (12mm,-0mm) rectangle (15mm,-3mm);
\fill[black!50] (15mm,-0mm) rectangle (18mm,-3mm);
\fill[black!0] (18mm,-0mm) rectangle (21mm,-3mm);
\fill[black!25] (21mm,-0mm) rectangle (24mm,-3mm);
\fill[black!0] (24mm,-0mm) rectangle (27mm,-3mm);
\fill[black!100] (27mm,-0mm) rectangle (30mm,-3mm);
\fill[black!0] (30mm,-0mm) rectangle (33mm,-3mm);
\fill[black!0] (33mm,-0mm) rectangle (36mm,-3mm);
\fill[black!0] (36mm,-0mm) rectangle (39mm,-3mm);
\fill[black!0] (0mm,-3mm) rectangle (3mm,-6mm);
\fill[black!0] (3mm,-3mm) rectangle (6mm,-6mm);
\fill[black!0] (6mm,-3mm) rectangle (9mm,-6mm);
\fill[black!0] (9mm,-3mm) rectangle (12mm,-6mm);
\fill[black!0] (12mm,-3mm) rectangle (15mm,-6mm);
\fill[black!50] (15mm,-3mm) rectangle (18mm,-6mm);
\fill[black!0] (18mm,-3mm) rectangle (21mm,-6mm);
\fill[black!0] (21mm,-3mm) rectangle (24mm,-6mm);
\fill[black!25] (24mm,-3mm) rectangle (27mm,-6mm);
\fill[black!100] (27mm,-3mm) rectangle (30mm,-6mm);
\fill[black!100] (30mm,-3mm) rectangle (33mm,-6mm);
\fill[black!100] (33mm,-3mm) rectangle (36mm,-6mm);
\fill[black!0] (36mm,-3mm) rectangle (39mm,-6mm);
\fill[black!0] (0mm,-6mm) rectangle (3mm,-9mm);
\fill[black!0] (3mm,-6mm) rectangle (6mm,-9mm);
\fill[black!0] (6mm,-6mm) rectangle (9mm,-9mm);
\fill[black!0] (9mm,-6mm) rectangle (12mm,-9mm);
\fill[black!0] (12mm,-6mm) rectangle (15mm,-9mm);
\fill[black!0] (15mm,-6mm) rectangle (18mm,-9mm);
\fill[black!25] (18mm,-6mm) rectangle (21mm,-9mm);
\fill[black!25] (21mm,-6mm) rectangle (24mm,-9mm);
\fill[black!0] (24mm,-6mm) rectangle (27mm,-9mm);
\fill[black!0] (27mm,-6mm) rectangle (30mm,-9mm);
\fill[black!100] (30mm,-6mm) rectangle (33mm,-9mm);
\fill[black!0] (33mm,-6mm) rectangle (36mm,-9mm);
\fill[black!0] (36mm,-6mm) rectangle (39mm,-9mm);
\fill[black!0] (0mm,-9mm) rectangle (3mm,-12mm);
\fill[black!0] (3mm,-9mm) rectangle (6mm,-12mm);
\fill[black!0] (6mm,-9mm) rectangle (9mm,-12mm);
\fill[black!0] (9mm,-9mm) rectangle (12mm,-12mm);
\fill[black!0] (12mm,-9mm) rectangle (15mm,-12mm);
\fill[black!0] (15mm,-9mm) rectangle (18mm,-12mm);
\fill[black!25] (18mm,-9mm) rectangle (21mm,-12mm);
\fill[black!0] (21mm,-9mm) rectangle (24mm,-12mm);
\fill[black!25] (24mm,-9mm) rectangle (27mm,-12mm);
\fill[black!0] (27mm,-9mm) rectangle (30mm,-12mm);
\fill[black!0] (30mm,-9mm) rectangle (33mm,-12mm);
\fill[black!100] (33mm,-9mm) rectangle (36mm,-12mm);
\fill[black!100] (36mm,-9mm) rectangle (39mm,-12mm);
\fill[black!0] (0mm,-12mm) rectangle (3mm,-15mm);
\fill[black!0] (3mm,-12mm) rectangle (6mm,-15mm);
\fill[black!0] (6mm,-12mm) rectangle (9mm,-15mm);
\fill[black!0] (9mm,-12mm) rectangle (12mm,-15mm);
\fill[black!0] (12mm,-12mm) rectangle (15mm,-15mm);
\fill[black!0] (15mm,-12mm) rectangle (18mm,-15mm);
\fill[black!25] (18mm,-12mm) rectangle (21mm,-15mm);
\fill[black!0] (21mm,-12mm) rectangle (24mm,-15mm);
\fill[black!0] (24mm,-12mm) rectangle (27mm,-15mm);
\fill[black!0] (27mm,-12mm) rectangle (30mm,-15mm);
\fill[black!0] (30mm,-12mm) rectangle (33mm,-15mm);
\fill[black!0] (33mm,-12mm) rectangle (36mm,-15mm);
\fill[black!100] (36mm,-12mm) rectangle (39mm,-15mm);
\fill[black!50] (0mm,-15mm) rectangle (3mm,-18mm);
\fill[black!50] (3mm,-15mm) rectangle (6mm,-18mm);
\fill[black!0] (6mm,-15mm) rectangle (9mm,-18mm);
\fill[black!0] (9mm,-15mm) rectangle (12mm,-18mm);
\fill[black!0] (12mm,-15mm) rectangle (15mm,-18mm);
\fill[black!0] (15mm,-15mm) rectangle (18mm,-18mm);
\fill[black!0] (18mm,-15mm) rectangle (21mm,-18mm);
\fill[black!0] (21mm,-15mm) rectangle (24mm,-18mm);
\fill[black!0] (24mm,-15mm) rectangle (27mm,-18mm);
\fill[black!100] (27mm,-15mm) rectangle (30mm,-18mm);
\fill[black!0] (30mm,-15mm) rectangle (33mm,-18mm);
\fill[black!0] (33mm,-15mm) rectangle (36mm,-18mm);
\fill[black!0] (36mm,-15mm) rectangle (39mm,-18mm);
\fill[black!0] (0mm,-18mm) rectangle (3mm,-21mm);
\fill[black!0] (3mm,-18mm) rectangle (6mm,-21mm);
\fill[black!25] (6mm,-18mm) rectangle (9mm,-21mm);
\fill[black!25] (9mm,-18mm) rectangle (12mm,-21mm);
\fill[black!25] (12mm,-18mm) rectangle (15mm,-21mm);
\fill[black!0] (15mm,-18mm) rectangle (18mm,-21mm);
\fill[black!0] (18mm,-18mm) rectangle (21mm,-21mm);
\fill[black!0] (21mm,-18mm) rectangle (24mm,-21mm);
\fill[black!0] (24mm,-18mm) rectangle (27mm,-21mm);
\fill[black!0] (27mm,-18mm) rectangle (30mm,-21mm);
\fill[black!25] (30mm,-18mm) rectangle (33mm,-21mm);
\fill[black!25] (33mm,-18mm) rectangle (36mm,-21mm);
\fill[black!25] (36mm,-18mm) rectangle (39mm,-21mm);
\fill[black!25] (0mm,-21mm) rectangle (3mm,-24mm);
\fill[black!0] (3mm,-21mm) rectangle (6mm,-24mm);
\fill[black!25] (6mm,-21mm) rectangle (9mm,-24mm);
\fill[black!0] (9mm,-21mm) rectangle (12mm,-24mm);
\fill[black!0] (12mm,-21mm) rectangle (15mm,-24mm);
\fill[black!0] (15mm,-21mm) rectangle (18mm,-24mm);
\fill[black!0] (18mm,-21mm) rectangle (21mm,-24mm);
\fill[black!0] (21mm,-21mm) rectangle (24mm,-24mm);
\fill[black!0] (24mm,-21mm) rectangle (27mm,-24mm);
\fill[black!25] (27mm,-21mm) rectangle (30mm,-24mm);
\fill[black!25] (30mm,-21mm) rectangle (33mm,-24mm);
\fill[black!0] (33mm,-21mm) rectangle (36mm,-24mm);
\fill[black!0] (36mm,-21mm) rectangle (39mm,-24mm);
\fill[black!0] (0mm,-24mm) rectangle (3mm,-27mm);
\fill[black!25] (3mm,-24mm) rectangle (6mm,-27mm);
\fill[black!0] (6mm,-24mm) rectangle (9mm,-27mm);
\fill[black!25] (9mm,-24mm) rectangle (12mm,-27mm);
\fill[black!0] (12mm,-24mm) rectangle (15mm,-27mm);
\fill[black!0] (15mm,-24mm) rectangle (18mm,-27mm);
\fill[black!0] (18mm,-24mm) rectangle (21mm,-27mm);
\fill[black!0] (21mm,-24mm) rectangle (24mm,-27mm);
\fill[black!0] (24mm,-24mm) rectangle (27mm,-27mm);
\fill[black!0] (27mm,-24mm) rectangle (30mm,-27mm);
\fill[black!0] (30mm,-24mm) rectangle (33mm,-27mm);
\fill[black!25] (33mm,-24mm) rectangle (36mm,-27mm);
\fill[black!25] (36mm,-24mm) rectangle (39mm,-27mm);
\fill[black!100] (0mm,-27mm) rectangle (3mm,-30mm);
\fill[black!100] (3mm,-27mm) rectangle (6mm,-30mm);
\fill[black!0] (6mm,-27mm) rectangle (9mm,-30mm);
\fill[black!0] (9mm,-27mm) rectangle (12mm,-30mm);
\fill[black!0] (12mm,-27mm) rectangle (15mm,-30mm);
\fill[black!100] (15mm,-27mm) rectangle (18mm,-30mm);
\fill[black!0] (18mm,-27mm) rectangle (21mm,-30mm);
\fill[black!25] (21mm,-27mm) rectangle (24mm,-30mm);
\fill[black!0] (24mm,-27mm) rectangle (27mm,-30mm);
\fill[black!0] (27mm,-27mm) rectangle (30mm,-30mm);
\fill[black!50] (30mm,-27mm) rectangle (33mm,-30mm);
\fill[black!0] (33mm,-27mm) rectangle (36mm,-30mm);
\fill[black!0] (36mm,-27mm) rectangle (39mm,-30mm);
\fill[black!0] (0mm,-30mm) rectangle (3mm,-33mm);
\fill[black!100] (3mm,-30mm) rectangle (6mm,-33mm);
\fill[black!100] (6mm,-30mm) rectangle (9mm,-33mm);
\fill[black!0] (9mm,-30mm) rectangle (12mm,-33mm);
\fill[black!0] (12mm,-30mm) rectangle (15mm,-33mm);
\fill[black!0] (15mm,-30mm) rectangle (18mm,-33mm);
\fill[black!25] (18mm,-30mm) rectangle (21mm,-33mm);
\fill[black!25] (21mm,-30mm) rectangle (24mm,-33mm);
\fill[black!0] (24mm,-30mm) rectangle (27mm,-33mm);
\fill[black!50] (27mm,-30mm) rectangle (30mm,-33mm);
\fill[black!0] (30mm,-30mm) rectangle (33mm,-33mm);
\fill[black!50] (33mm,-30mm) rectangle (36mm,-33mm);
\fill[black!0] (36mm,-30mm) rectangle (39mm,-33mm);
\fill[black!0] (0mm,-33mm) rectangle (3mm,-36mm);
\fill[black!100] (3mm,-33mm) rectangle (6mm,-36mm);
\fill[black!0] (6mm,-33mm) rectangle (9mm,-36mm);
\fill[black!100] (9mm,-33mm) rectangle (12mm,-36mm);
\fill[black!0] (12mm,-33mm) rectangle (15mm,-36mm);
\fill[black!0] (15mm,-33mm) rectangle (18mm,-36mm);
\fill[black!25] (18mm,-33mm) rectangle (21mm,-36mm);
\fill[black!0] (21mm,-33mm) rectangle (24mm,-36mm);
\fill[black!25] (24mm,-33mm) rectangle (27mm,-36mm);
\fill[black!0] (27mm,-33mm) rectangle (30mm,-36mm);
\fill[black!50] (30mm,-33mm) rectangle (33mm,-36mm);
\fill[black!0] (33mm,-33mm) rectangle (36mm,-36mm);
\fill[black!50] (36mm,-33mm) rectangle (39mm,-36mm);
\fill[black!0] (0mm,-36mm) rectangle (3mm,-39mm);
\fill[black!0] (3mm,-36mm) rectangle (6mm,-39mm);
\fill[black!0] (6mm,-36mm) rectangle (9mm,-39mm);
\fill[black!100] (9mm,-36mm) rectangle (12mm,-39mm);
\fill[black!100] (12mm,-36mm) rectangle (15mm,-39mm);
\fill[black!0] (15mm,-36mm) rectangle (18mm,-39mm);
\fill[black!25] (18mm,-36mm) rectangle (21mm,-39mm);
\fill[black!0] (21mm,-36mm) rectangle (24mm,-39mm);
\fill[black!25] (24mm,-36mm) rectangle (27mm,-39mm);
\fill[black!0] (27mm,-36mm) rectangle (30mm,-39mm);
\fill[black!0] (30mm,-36mm) rectangle (33mm,-39mm);
\fill[black!50] (33mm,-36mm) rectangle (36mm,-39mm);
\fill[black!0] (36mm,-36mm) rectangle (39mm,-39mm);
            \draw (current bounding box.north east) rectangle (current bounding box.south west);
        \end{tikzpicture}
        \label{fig:dcsh-pre}
    }
    \quad\quad\quad\quad\quad\quad
    \subfloat[After]{
        \begin{tikzpicture}
            \fill[black!0] (0mm,-0mm) rectangle (3mm,-3mm);
\fill[black!100] (3mm,-0mm) rectangle (6mm,-3mm);
\fill[black!50] (6mm,-0mm) rectangle (9mm,-3mm);
\fill[black!25] (9mm,-0mm) rectangle (12mm,-3mm);
\fill[black!0] (12mm,-0mm) rectangle (15mm,-3mm);
\fill[black!0] (15mm,-0mm) rectangle (18mm,-3mm);
\fill[black!0] (18mm,-0mm) rectangle (21mm,-3mm);
\fill[black!0] (21mm,-0mm) rectangle (24mm,-3mm);
\fill[black!0] (24mm,-0mm) rectangle (27mm,-3mm);
\fill[black!0] (27mm,-0mm) rectangle (30mm,-3mm);
\fill[black!0] (30mm,-0mm) rectangle (33mm,-3mm);
\fill[black!0] (33mm,-0mm) rectangle (36mm,-3mm);
\fill[black!0] (36mm,-0mm) rectangle (39mm,-3mm);
\fill[black!100] (0mm,-3mm) rectangle (3mm,-6mm);
\fill[black!0] (3mm,-3mm) rectangle (6mm,-6mm);
\fill[black!100] (6mm,-3mm) rectangle (9mm,-6mm);
\fill[black!25] (9mm,-3mm) rectangle (12mm,-6mm);
\fill[black!100] (12mm,-3mm) rectangle (15mm,-6mm);
\fill[black!50] (15mm,-3mm) rectangle (18mm,-6mm);
\fill[black!0] (18mm,-3mm) rectangle (21mm,-6mm);
\fill[black!0] (21mm,-3mm) rectangle (24mm,-6mm);
\fill[black!0] (24mm,-3mm) rectangle (27mm,-6mm);
\fill[black!0] (27mm,-3mm) rectangle (30mm,-6mm);
\fill[black!0] (30mm,-3mm) rectangle (33mm,-6mm);
\fill[black!0] (33mm,-3mm) rectangle (36mm,-6mm);
\fill[black!0] (36mm,-3mm) rectangle (39mm,-6mm);
\fill[black!50] (0mm,-6mm) rectangle (3mm,-9mm);
\fill[black!100] (3mm,-6mm) rectangle (6mm,-9mm);
\fill[black!0] (6mm,-6mm) rectangle (9mm,-9mm);
\fill[black!0] (9mm,-6mm) rectangle (12mm,-9mm);
\fill[black!50] (12mm,-6mm) rectangle (15mm,-9mm);
\fill[black!0] (15mm,-6mm) rectangle (18mm,-9mm);
\fill[black!0] (18mm,-6mm) rectangle (21mm,-9mm);
\fill[black!0] (21mm,-6mm) rectangle (24mm,-9mm);
\fill[black!0] (24mm,-6mm) rectangle (27mm,-9mm);
\fill[black!0] (27mm,-6mm) rectangle (30mm,-9mm);
\fill[black!0] (30mm,-6mm) rectangle (33mm,-9mm);
\fill[black!0] (33mm,-6mm) rectangle (36mm,-9mm);
\fill[black!0] (36mm,-6mm) rectangle (39mm,-9mm);
\fill[black!25] (0mm,-9mm) rectangle (3mm,-12mm);
\fill[black!25] (3mm,-9mm) rectangle (6mm,-12mm);
\fill[black!0] (6mm,-9mm) rectangle (9mm,-12mm);
\fill[black!0] (9mm,-9mm) rectangle (12mm,-12mm);
\fill[black!0] (12mm,-9mm) rectangle (15mm,-12mm);
\fill[black!25] (15mm,-9mm) rectangle (18mm,-12mm);
\fill[black!25] (18mm,-9mm) rectangle (21mm,-12mm);
\fill[black!0] (21mm,-9mm) rectangle (24mm,-12mm);
\fill[black!0] (24mm,-9mm) rectangle (27mm,-12mm);
\fill[black!0] (27mm,-9mm) rectangle (30mm,-12mm);
\fill[black!0] (30mm,-9mm) rectangle (33mm,-12mm);
\fill[black!0] (33mm,-9mm) rectangle (36mm,-12mm);
\fill[black!0] (36mm,-9mm) rectangle (39mm,-12mm);
\fill[black!0] (0mm,-12mm) rectangle (3mm,-15mm);
\fill[black!100] (3mm,-12mm) rectangle (6mm,-15mm);
\fill[black!50] (6mm,-12mm) rectangle (9mm,-15mm);
\fill[black!0] (9mm,-12mm) rectangle (12mm,-15mm);
\fill[black!0] (12mm,-12mm) rectangle (15mm,-15mm);
\fill[black!100] (15mm,-12mm) rectangle (18mm,-15mm);
\fill[black!0] (18mm,-12mm) rectangle (21mm,-15mm);
\fill[black!100] (21mm,-12mm) rectangle (24mm,-15mm);
\fill[black!25] (24mm,-12mm) rectangle (27mm,-15mm);
\fill[black!0] (27mm,-12mm) rectangle (30mm,-15mm);
\fill[black!0] (30mm,-12mm) rectangle (33mm,-15mm);
\fill[black!0] (33mm,-12mm) rectangle (36mm,-15mm);
\fill[black!0] (36mm,-12mm) rectangle (39mm,-15mm);
\fill[black!0] (0mm,-15mm) rectangle (3mm,-18mm);
\fill[black!50] (3mm,-15mm) rectangle (6mm,-18mm);
\fill[black!0] (6mm,-15mm) rectangle (9mm,-18mm);
\fill[black!25] (9mm,-15mm) rectangle (12mm,-18mm);
\fill[black!100] (12mm,-15mm) rectangle (15mm,-18mm);
\fill[black!0] (15mm,-15mm) rectangle (18mm,-18mm);
\fill[black!100] (18mm,-15mm) rectangle (21mm,-18mm);
\fill[black!50] (21mm,-15mm) rectangle (24mm,-18mm);
\fill[black!0] (24mm,-15mm) rectangle (27mm,-18mm);
\fill[black!25] (27mm,-15mm) rectangle (30mm,-18mm);
\fill[black!0] (30mm,-15mm) rectangle (33mm,-18mm);
\fill[black!0] (33mm,-15mm) rectangle (36mm,-18mm);
\fill[black!0] (36mm,-15mm) rectangle (39mm,-18mm);
\fill[black!0] (0mm,-18mm) rectangle (3mm,-21mm);
\fill[black!0] (3mm,-18mm) rectangle (6mm,-21mm);
\fill[black!0] (6mm,-18mm) rectangle (9mm,-21mm);
\fill[black!25] (9mm,-18mm) rectangle (12mm,-21mm);
\fill[black!0] (12mm,-18mm) rectangle (15mm,-21mm);
\fill[black!100] (15mm,-18mm) rectangle (18mm,-21mm);
\fill[black!0] (18mm,-18mm) rectangle (21mm,-21mm);
\fill[black!0] (21mm,-18mm) rectangle (24mm,-21mm);
\fill[black!0] (24mm,-18mm) rectangle (27mm,-21mm);
\fill[black!25] (27mm,-18mm) rectangle (30mm,-21mm);
\fill[black!0] (30mm,-18mm) rectangle (33mm,-21mm);
\fill[black!0] (33mm,-18mm) rectangle (36mm,-21mm);
\fill[black!0] (36mm,-18mm) rectangle (39mm,-21mm);
\fill[black!0] (0mm,-21mm) rectangle (3mm,-24mm);
\fill[black!0] (3mm,-21mm) rectangle (6mm,-24mm);
\fill[black!0] (6mm,-21mm) rectangle (9mm,-24mm);
\fill[black!0] (9mm,-21mm) rectangle (12mm,-24mm);
\fill[black!100] (12mm,-21mm) rectangle (15mm,-24mm);
\fill[black!50] (15mm,-21mm) rectangle (18mm,-24mm);
\fill[black!0] (18mm,-21mm) rectangle (21mm,-24mm);
\fill[black!0] (21mm,-21mm) rectangle (24mm,-24mm);
\fill[black!25] (24mm,-21mm) rectangle (27mm,-24mm);
\fill[black!25] (27mm,-21mm) rectangle (30mm,-24mm);
\fill[black!100] (30mm,-21mm) rectangle (33mm,-24mm);
\fill[black!50] (33mm,-21mm) rectangle (36mm,-24mm);
\fill[black!0] (36mm,-21mm) rectangle (39mm,-24mm);
\fill[black!0] (0mm,-24mm) rectangle (3mm,-27mm);
\fill[black!0] (3mm,-24mm) rectangle (6mm,-27mm);
\fill[black!0] (6mm,-24mm) rectangle (9mm,-27mm);
\fill[black!0] (9mm,-24mm) rectangle (12mm,-27mm);
\fill[black!25] (12mm,-24mm) rectangle (15mm,-27mm);
\fill[black!0] (15mm,-24mm) rectangle (18mm,-27mm);
\fill[black!0] (18mm,-24mm) rectangle (21mm,-27mm);
\fill[black!25] (21mm,-24mm) rectangle (24mm,-27mm);
\fill[black!0] (24mm,-24mm) rectangle (27mm,-27mm);
\fill[black!0] (27mm,-24mm) rectangle (30mm,-27mm);
\fill[black!25] (30mm,-24mm) rectangle (33mm,-27mm);
\fill[black!25] (33mm,-24mm) rectangle (36mm,-27mm);
\fill[black!0] (36mm,-24mm) rectangle (39mm,-27mm);
\fill[black!0] (0mm,-27mm) rectangle (3mm,-30mm);
\fill[black!0] (3mm,-27mm) rectangle (6mm,-30mm);
\fill[black!0] (6mm,-27mm) rectangle (9mm,-30mm);
\fill[black!0] (9mm,-27mm) rectangle (12mm,-30mm);
\fill[black!0] (12mm,-27mm) rectangle (15mm,-30mm);
\fill[black!25] (15mm,-27mm) rectangle (18mm,-30mm);
\fill[black!25] (18mm,-27mm) rectangle (21mm,-30mm);
\fill[black!25] (21mm,-27mm) rectangle (24mm,-30mm);
\fill[black!0] (24mm,-27mm) rectangle (27mm,-30mm);
\fill[black!0] (27mm,-27mm) rectangle (30mm,-30mm);
\fill[black!25] (30mm,-27mm) rectangle (33mm,-30mm);
\fill[black!25] (33mm,-27mm) rectangle (36mm,-30mm);
\fill[black!25] (36mm,-27mm) rectangle (39mm,-30mm);
\fill[black!0] (0mm,-30mm) rectangle (3mm,-33mm);
\fill[black!0] (3mm,-30mm) rectangle (6mm,-33mm);
\fill[black!0] (6mm,-30mm) rectangle (9mm,-33mm);
\fill[black!0] (9mm,-30mm) rectangle (12mm,-33mm);
\fill[black!0] (12mm,-30mm) rectangle (15mm,-33mm);
\fill[black!0] (15mm,-30mm) rectangle (18mm,-33mm);
\fill[black!0] (18mm,-30mm) rectangle (21mm,-33mm);
\fill[black!100] (21mm,-30mm) rectangle (24mm,-33mm);
\fill[black!25] (24mm,-30mm) rectangle (27mm,-33mm);
\fill[black!25] (27mm,-30mm) rectangle (30mm,-33mm);
\fill[black!0] (30mm,-30mm) rectangle (33mm,-33mm);
\fill[black!100] (33mm,-30mm) rectangle (36mm,-33mm);
\fill[black!0] (36mm,-30mm) rectangle (39mm,-33mm);
\fill[black!0] (0mm,-33mm) rectangle (3mm,-36mm);
\fill[black!0] (3mm,-33mm) rectangle (6mm,-36mm);
\fill[black!0] (6mm,-33mm) rectangle (9mm,-36mm);
\fill[black!0] (9mm,-33mm) rectangle (12mm,-36mm);
\fill[black!0] (12mm,-33mm) rectangle (15mm,-36mm);
\fill[black!0] (15mm,-33mm) rectangle (18mm,-36mm);
\fill[black!0] (18mm,-33mm) rectangle (21mm,-36mm);
\fill[black!50] (21mm,-33mm) rectangle (24mm,-36mm);
\fill[black!25] (24mm,-33mm) rectangle (27mm,-36mm);
\fill[black!25] (27mm,-33mm) rectangle (30mm,-36mm);
\fill[black!100] (30mm,-33mm) rectangle (33mm,-36mm);
\fill[black!0] (33mm,-33mm) rectangle (36mm,-36mm);
\fill[black!100] (36mm,-33mm) rectangle (39mm,-36mm);
\fill[black!0] (0mm,-36mm) rectangle (3mm,-39mm);
\fill[black!0] (3mm,-36mm) rectangle (6mm,-39mm);
\fill[black!0] (6mm,-36mm) rectangle (9mm,-39mm);
\fill[black!0] (9mm,-36mm) rectangle (12mm,-39mm);
\fill[black!0] (12mm,-36mm) rectangle (15mm,-39mm);
\fill[black!0] (15mm,-36mm) rectangle (18mm,-39mm);
\fill[black!0] (18mm,-36mm) rectangle (21mm,-39mm);
\fill[black!0] (21mm,-36mm) rectangle (24mm,-39mm);
\fill[black!0] (24mm,-36mm) rectangle (27mm,-39mm);
\fill[black!25] (27mm,-36mm) rectangle (30mm,-39mm);
\fill[black!0] (30mm,-36mm) rectangle (33mm,-39mm);
\fill[black!100] (33mm,-36mm) rectangle (36mm,-39mm);
\fill[black!0] (36mm,-36mm) rectangle (39mm,-39mm);
            \draw (current bounding box.north east) rectangle (current bounding box.south west);
        \end{tikzpicture}
        \label{fig:dcsh-post}
    }
    \caption{Example of a DSM, before and after applying DCSH.}
    \label{fig:dcsh}
\end{figure}
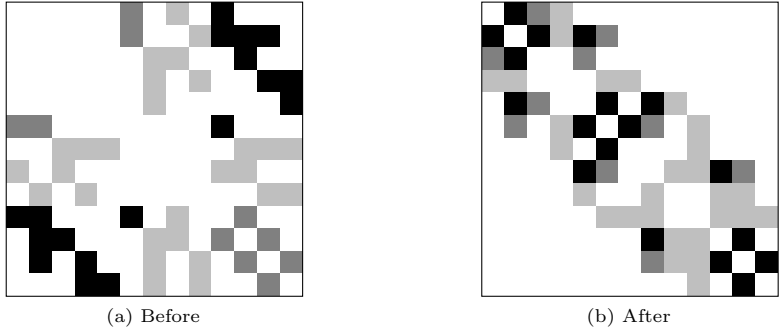

Compared to the DCSH algorithm from \cite{Lousberg2020}, the CIF implementation can moreover deal with multiple unconnected partitions within the adjacency graph, by ordering each of them separately.
The partitions are sorted in descending order based on their size, such that the variables corresponding to unconnected nodes (single-variable partitions) still appear at the end of the resulting variable order, as in~\cite{Lousberg2020}.

An example of the effects of applying DCSH is shown in Figure~\ref{fig:dcsh}.
It shows both the DSM before applying DCSH (\ref{fig:dcsh-pre}) and the one after applying DCSH (\ref{fig:dcsh-post}).
From top to bottom, and left to right, the variables are shown, either in the original model order (before ordering) or in the computed order (after ordering).
The shading of the cell indicates the weight between the variables (the darker the cell, the larger the weight).

ESCET v0.8 uses only the FORCE algorithm~\cite{Aloul2003}, followed by the classic sliding window algorithm, to automatically order the variables.
ESCET v4.0 first uses the DCSH algorithm, before still applying the FORCE and sliding window algorithms.
Lousberg et al.~have shown that applying FORCE after DCSH produces better results than only applying one of them~\cite{Lousberg2020}.
The sliding window algorithm further locally optimizes the variable order.

\subsection{Linearized variable ordering relations for BDDs}
\label{sec:linearized-var-order-relations}

Heuristic variable ordering algorithms for BDDs, like DCSH, require as input some information about the relations between the variables that are to be ordered, for instance in the form of a DSM or an adjacency graph.
The variable relations are also determined using heuristics, and using different variable relation heuristics may lead to different variable orders produced by variable ordering algorithms, which affects synthesis performance.
Synthesis performance can therefore be improved by choosing a variable relations heuristic in such a way that it fits well with the reachability computations (see Algorithm \ref{alg:reach}), a key ingredient of the symbolic synthesis algorithm (see Algorithm \ref{alg:synth}), and especially with the transition relations on which these reachability computations operate.

The symbolic synthesis algorithm in the ESCET toolkit linearizes the CIF model to eliminate the parallel composition between the different EFAs, resulting in a single EFA with a single location and only self-loop edges, which is then encoded as a SEFA (as previously explained in Section~\ref{sec:cif-synth-algo}).
Since the transition relations -- the SEFA edges -- are directly derived from the linearized edges, it is beneficial to also derive the variable relations from the linearized edges.

Lousberg et al.~\cite{Lousberg2020} propose to count the number of times that two variables occur together in the different linearized edges, and use that `weight' as the relation between two variables.
That is, in the DSM the cells are filled with these weights, and in the adjacency graph the edges are labeled with the weights.
By using this particular variable relation, variable ordering algorithms places all variables that occur more frequently together in transition relations more closely together in the variable order.
This allows the transition relations to be applied more efficiently during synthesis.

For the linearized edges of Figure~\ref{fig:linearization-post}, the variables that occur in the linearized edges are:
\begin{align*}
    e_1 &: l_1, v       & e_6~\, &: l_1, l_2, x \\
    e_2 &: l_1, x       & e_7~\, &: l_1, l_2, x \\
    e_3 &: l_1, x       & e_8~\, &: l_1, l_2, x \\
    e_4 &: l_2, v, x, y & e_9~\, &: l_1, v, x \\
    e_5 &: l_1, l_2, x  & e_{10} &: l_2
\end{align*}
        
The corresponding linearized variable relations are shown in Figure~\ref{fig:var-rel-dsm} as a DSM.
For instance, variables $l_1$ and $x$ occur together in seven linearized edges.
Self-relations of variables are excluded, and thus the top-left to bottom-right diagonal always only contains entries with value zero.

\begin{figure}[t]
    \centering
    \begin{tikzpicture}
        \input{images/var-rel-dsm.tex}
    \end{tikzpicture}
    \caption{Example DSM with linearized variable relations derived from linearized edges.}
    \label{fig:var-rel-dsm}
\end{figure}

ESCET v0.8 does not have these linearized variable ordering relations, but instead uses more complex custom relations.
ESCET v4.0 uses the linearized variable ordering relations for the FORCE and sliding window algorithms, and the legacy relations from v0.8 for the DCSH algorithm.
This combination was found to give the best results.

\subsection{Per-event edge granularity}
\label{sec:per-event-edge-granularity}

The transition relations that are used during synthesis are derived from the linearized edges.
It is possible to use each linearized edge as a separate transition relation (as shown in Section~\ref{sec:sefa}), applying them one by one during reachability computations (see Algorithm~\ref{alg:reach}).
SEFA edges and transition relations then match one-on-one.
It is also possible to combine multiple transition relations into a single transition relation~\cite{Burch1994}.
Two SEFA edges $\cramped{(g_1,r_1,\sigma,u_1),(g_2,r_2,\sigma,u_2)\in E}$ can be combined to a single edge $(g, r, \sigma, u)$ as follows:
\vspace{-1mm}
\begin{align*}
    g &= g_1 \lor g_2 \\
    r &= (g_1 \land r_1) \lor (g_2 \land r_2) \\
    u &= (g_1 \land u'_1) \lor (g_2 \land u'_2)
\end{align*}
\vspace{-1mm}
Update $u'_1$ is obtained from $u_1$ by adding $x^+ \!= x$ for each variable $x$ that is assigned in update $u_2$ but not in $u_1$, and similarly update $u'_2$ is obtained from $u_2$.

As an example, consider the following two SEFA edges, and their combination: 
\begin{align*}
    \text{Edge 1:}&~(x\leq 4, \textit{true}, e, y^+\!=y+1) \\
    \text{Edge 2:}&~(x\geq 4, \textit{true}, e, z^+\!=z+1) \\
    \text{Combined:}&~(\textit{true}, \textit{true}, e, \\
            &\quad (x\leq 4 \land y^+\!=y+1 \land z^+\!=z) \,\lor \\
            &\quad (x\geq 4 \land y^+\!=y \land z^+\!=z+1) \\
            &~)
\end{align*}

Combining similar transition relations leads to fewer transition relations that need to be applied during reachability computations.
The combined transition relation may be simpler than the original ones, as in the example above the guards $x\leq 4$ and $x\geq 4$ get combined to \emph{true}.
This kind of reduction is not uncommon, but there are also cases where the combination is more complex than the original guards.
Similarly, update predicates may also get reduced, or become more complex.
In the example above, they get more complex, as they mostly concern different variables (one increments $y$, the other increments $z$).
Combining edges is often a trade-off between the number of transition relations and their size.

In ESCET v0.8, CIF's symbolic synthesis tool uses the per-linearized-edge transition relations.
In ESCET v4.0, it combines all linearized transition relations, per event.
This level of granularity is more combined than the per-linearized-edge transition relations, but less so than combining all transitions relations for all events into a single transition relation (which is not allowed by our definition).
Using a single transition relation for all events often leads to much larger transition relations and intermediate results during synthesis, compared to using per-event transition relations, as shown by Fei et al.~\cite{Fei2014}.

Per-event transition relation granularity makes sense in our case, since linearized edges with the same event will often (at least in part) have similar guards and updates, as they are constructed by combining the edges of the original automata from the CIF model in all different combinations.
We saw an example earlier in Figure~\ref{fig:linearization-post}, where the four linearized edges for event \texttt{decide} have similar guards and updates.
Using per-event transition relation granularity, in practice, the number of transition relations is reduced, while their size does not increase too much.

\subsection{Efficient edge application}
\label{sec:efficient-edge-application}

Since applying edges is a key operation in synthesis, applying edges more efficiently can significantly improve the synthesis performance.
This can, among others, be achieved by removing redundant operations and by using compounded operations that combine multiple separate operations.

In ESCET v4.0, the synthesis algorithm uses the \emph{relnext(Intersection)} and \emph{relprev(Intersection)} BDD operations to efficiently compute SEFA transitions by applying SEFA edges (see Section~\ref{sec:edge-apply}).
Remember that $\textit{relnextIntersection}(e, p, R)$ computes for an edge $\cramped{e=(g,r,\sigma,u)}$, predicate $p$ and restriction $R$, the predicate $(\exists_{\textit{vars}(t)}(p \land t))[V^+ := V] \land R$, with $t = g \land \lnot\, r \land u$ being precomputed.

In ESCET v0.8, applying edges is less efficient.
It does not use \emph{relnextIntersection}.
It computes the predicate $(\exists_{V}(p \land g \land \lnot\, r \land u))[V^+ := V] \land R$, with $g \land u \land R$ being precomputed.
These differences have several performance implications.

First of all, transition relation $t$ does not change during synthesis, and thus needs to be computed only once.
Contrarily, $g \land u \land R$ contains $R$, which does change during synthesis; in Algorithm~\ref{alg:synth}, controlled-behavior predicate $C$ is passed for $R$, which is constantly updated during the algorithm.
In v0.8, the precomputation is therefore done for each reachability computation, rather than once before all reachability computations, leading to more computations.

Secondly, in v4.0, the transition relation includes error predicate $r$.
$r$ is then applied together with guard $g$ and update $u$.
In v0.8, $r$ is applied using a separate conjunction operation.
This is less efficient, as it uses more separate operations that each have to go through the full BDDs and modify them.

Thirdly, in v0.8, the existential quantification and replacement operation are applied as separate BDD operations.
In v4.0, this is all part of the compounded operations \emph{relnext} and \emph{relprev}, which are more efficient.

And fourthly, v4.0 uses partial transition relations, for which a reduced existential quantification over a more limited set of variables can be used.
v0.8 uses full transition relations, with $x^+ \!= x$ in updates $u$ for each variable $x$ not assigned on the edge, leading to larger predicates and more work.
v0.8 further uses an existential quantification over the full variable set $V$, which also requires more work.

\subsection{Early fixed-point detection}
\label{sec:early-fixed-point-detection}

Symbolic supervisory controller synthesis involves several fixed-point computations (see Algorithm \ref{alg:synth}), where edges (transition relations) are repeatedly applied to a certain predicate representing a part of the state space, and this is repeated until the state space does not grow any further (see Algorithm \ref{alg:reach}).
If after applying certain edges it becomes clear that no more states will be reached, then the reachability computation is completed (line 4 in Algorithm \ref{alg:reach}).

A naive approach for computing reachability, as shown in Algorithm \ref{alg:reach}, is to apply each edge once, and subsequently apply all of them again if any of them led to reaching more states, and to keep repeating this for as long as more states are reached.
If in a certain iteration of applying the edges no new states are reached, the reachability computation terminates.

An example is shown in Figure~\ref{fig:reach-fixed-point-naive}.
There are six edges, that are applied one by one, each iteration.
In the first iteration, four edges reach new states (indicated by `\checkmark'), and two do not (indicated by `$\times$').
Since at least one edge reached new states, a second iteration is needed.
In that second iteration, three edges reach new states, and thus a third iteration is needed.
In that third iteration, none of the edges reach new states, and thus the reachability computation has reached a fixed point.

\begin{figure}[t!]
    \centering
    \subfloat[Naive approach]{
        \begin{tabular}{cccc}
            \textbf{Edge} & \multicolumn{3}{c}{\textbf{Iteration}} \\
            ~ & \textbf{1} & \textbf{2} & \textbf{3} \\
            $e_1$ & \checkmark & $\times$   & $\times$ \\
            $e_2$ & $\times$   & \checkmark & $\times$ \\
            $e_3$ & \checkmark & \checkmark & $\times$ \\
            $e_4$ & $\times$   & \checkmark & $\times$ \\
            $e_5$ & \checkmark & $\times$   & $\times$ \\
            $e_6$ & \checkmark & $\times$   & $\times$ \\
        \end{tabular}
        \label{fig:reach-fixed-point-naive}
    }
    \quad\quad\quad\quad\quad\quad
    \subfloat[Stop earlier]{
        \begin{tabular}{cccc}
            \textbf{Edge} & \multicolumn{3}{c}{\textbf{Iteration}} \\
            ~ & \textbf{1} & \textbf{2} & \textbf{3} \\
            $e_1$ & \checkmark & $\times$   & $\times$ \\
            $e_2$ & $\times$   & \checkmark & $\times$ \\
            $e_3$ & \checkmark & \checkmark & $\times$ \\
            $e_4$ & $\times$   & \checkmark & $\times$ \\
            $e_5$ & \checkmark & $\times$   & -- \\
            $e_6$ & \checkmark & $\times$   & -- \\
        \end{tabular}
        \label{fig:reach-fixed-point-earlier}
    }
    \caption{Example of concluding earlier that a reachability fixed point is reached.}
    \label{fig:reach-fixed-point}
\end{figure}

Assuming there are $n$ edges that are ordered in some way, then it is also possible to stop earlier in the last iteration.
For instance, in the example, in the second iteration of applying the edges, more states are reached by applying $e_2$, $e_3$ and $e_4$, but not by $e_5$ and $e_6$ afterwards.
Once all $n$ edges have been applied once more, after the last edge that reached new states, the algorithm could already stop.
That is, in the third iteration, there is no need to apply the last two edges (indicated by `--'), as they would be applied the same way to the same predicate as in the previous iteration, and would thus again not reach any new states.
This is shown in Figure~\ref{fig:reach-fixed-point-earlier}.

In this example, by stopping earlier, two edges do not need to be applied.
In general, the number of edges that does not need to be applied in the last iteration may vary within the range [1,$n$].

ESCET v0.8 uses Algorithm~\ref{alg:reach}, while ESCET v4.0 uses the optimized version described in this section.

\subsection{Invariants: improved state plant invariants enforcement}
\label{sec:improved-state-plant-inv-enforce}

In CIF, the behavior of plant automata can be restricted using plant invariants (see Section~\ref{sec:cif-models}).
They are taken into account during synthesis by restricting the guards of edges (see Section~\ref{sec:linearized-model-to-sefa}).
In ESCET v4.0, the approach of Thuijsman et al.~\cite{Thuijsman2021} is used to optimize that, by forgoing updating the edge guard if it is not required.

In ESCET v0.8, a different approach is used to prevent the guards from being needlessly updated.
Instead of checking whether $s$ is already implied by $g \land p_p$, it simplifies $s$ with respect to $p_p$ to produce $s'$, and then always adds $s'$ to the guard by changing $g$ to $g \land s'$.
The simplification algorithm is based on Coudert and Madre's `restrict' function~\cite{Coudert1990}.

The simplification approach of Thuijsman et al.~\cite{Thuijsman2021} is superior, as it computes with certainty whether an additional restriction is needed, while the simplification is based on best-effort heuristics, and therefore does not offer the same guarantees.

\section{Benchmark models}
\label{sec:benchmarks}

To be able to perform performance measurements, suitable benchmark models are needed.
CIF ships with a set of benchmark models.
They consist of a mix of industrial and academic models that differ in size and complexity, making them suitable for performance evaluations.
We will use these benchmarks in Section \ref{sec:experiments} to measure the performance improvements described in Section \ref{sec:improvements}.
The models are shared under an open license, the MIT license, and can thus also be used by other researchers to benchmark their synthesis algorithms.

In this section, we introduce and describe CIF's set of benchmark models and provide various metrics for them.
The artifact that accompanies this paper includes both the models and the scripts to produce the metrics~\cite{ArtifactOfThisPaper}.
The benchmark models themselves are not new, and are thus not a contribution of this paper.

\subsection{Overview of benchmark models}

Table~\ref{tbl:benchmark-models} shows an overview of the benchmark models shipped with Eclipse ESCET v4.0 that we use for our experiments in Section~\ref{sec:experiments}.
For each benchmark, the table lists the model number, the short and full names of the model, references to publications with further information (if available), the size of the uncontrolled system state space (`US', in number of states), and the size of the controlled system state space (`CS', in number of states).
Although the state space sizes do not translate one-to-one to the sizes of the BDD representations of the models, nor to the complexity of the computations performed during synthesis, they still provide some indication of the complexity of the models.
For some benchmarks, the number of states in the controlled system is larger than that of the uncontrolled system.
While the requirements restrict the behavior of the uncontrolled system, in such cases the requirements themselves also introduce additional state to monitor the plants.

\begin{table*}
    \caption{Overview of the benchmark models.}
    \label{tbl:benchmark-models}
    \centering
    \small
    \setlength{\tabcolsep}{12pt}
    \begin{tabular}{ l l l l l l  }

    \textbf{Nr} &
    \textbf{Short name} &
    \textbf{Full name} &
    \textbf{Refs} &
    \textbf{Size (US)} &
    \textbf{Size (CS)} \\

    1 &
    adas &
    Advanced driver assistance system &
    \cite{Korssen2018} &
    $3.40 \cdot 10^{09}$ &
    $2.04 \cdot 10^{10}$ \\

    2 &
    agv &
    Automated guided vehicles &
    \cite{Wonham2019} &
    $3.07 \cdot 10^{03}$ &
    $4.41 \cdot 10^{03}$ \\

    3 &
    bcs-dynamic &
    Body comfort system (dynamic) &
    \cite{Beek2016,Thuijsman2023} &
    $6.16 \cdot 10^{20}$ &
    $4.23 \cdot 10^{20}$ \\

    4 &
    bcs-static &
    Body comfort system (static) &
    \cite{Beek2016,Thuijsman2023} &
    $3.18 \cdot 10^{14}$ &
    $1.97 \cdot 10^{14}$ \\

    5 &
    bridge &
    Bridge &
    \cite{Reijnen2018a} &
    $5.92 \cdot 10^{33}$ &
    $9.55 \cdot 10^{28}$ \\

    6 &
    cmt-v1 &
    Cat and mouse tower (v1, $n\!=\!3$, $k\!=\!2$) &
    \cite{Ma2008,Thuijsman2021} &
    $1.44 \cdot 10^{04}$ &
    $8.61 \cdot 10^{03}$ \\

    7 &
    cmt-v2 &
    Cat and mouse tower (v2, $n\!=\!3$, $k\!=\!1$) &
    \cite{Ma2008,Thuijsman2021} &
    $1.07 \cdot 10^{09}$ &
    $3.03 \cdot 10^{05}$ \\

    8 &
    cluster-tool &
    Cluster tool &
    \cite{Su2010} &
    $7.50 \cdot 10^{06}$ &
    $2.70 \cdot 10^{06}$ \\

    9 &
    dining-phils &
    Dining philosophers &
    \cite{Hoare1985} &
    $2.43 \cdot 10^{02}$ &
    $2.41 \cdot 10^{02}$ \\

    10 &
    festo &
    FESTO production line &
    \cite{Reijnen2018b} &
    $1.47 \cdot 10^{26}$ &
    $2.22 \cdot 10^{25}$ \\

    11 &
    litho-init &
    Lithography machine initialization &
    \cite{Vos2020,Thuijsman2021} &
    $7.21 \cdot 10^{09}$ &
    $2.46 \cdot 10^{09}$ \\

    12 &
    mri-pss-event &
    MRI scanner PSS (event-based) &
    \cite{Theunissen2014,Theunissen2015} &
    $3.60 \cdot 10^{03}$ &
    $3.09 \cdot 10^{04}$ \\

    13 &
    mri-pss-state &
    MRI scanner PSS (state-based) &
    \cite{Theunissen2015} &
    $1.44 \cdot 10^{04}$ &
    $2.79 \cdot 10^{04}$ \\

    14 &
    multi-agent-form &
    Multi-agent formation &
    \cite{Cai2015} &
    $1.00 \cdot 10^{03}$ &
    $3.04 \cdot 10^{02}$ \\

    15 &
    prod-cell &
    Production cell &
    \cite{Feng2009} &
    $3.76 \cdot 10^{08}$ &
    $1.15 \cdot 10^{08}$ \\

    16 &
    robo-swarm-aggr &
    Robotic swarm aggregation &
    \cite{Lopes2016} &
    $1.00 \cdot 10^{00}$ &
    $9.00 \cdot 10^{00}$ \\

    17 &
    robo-swarm-clus &
    Robotic swarm clustering &
    \cite{Lopes2016} &
    $1.00 \cdot 10^{00}$ &
    $1.60 \cdot 10^{01}$ \\

    18 &
    robo-swarm-form &
    Robotic swarm formation &
    \cite{Lopes2016} &
    $8.00 \cdot 10^{01}$ &
    $1.38 \cdot 10^{02}$ \\

    19 &
    robo-swarm-segr &
    Robotic swarm segregation &
    \cite{Lopes2016} &
    $6.40 \cdot 10^{01}$ &
    $1.28 \cdot 10^{02}$ \\

    20 &
    sudoku &
    Sudoku &
    ~ &
    $1.53 \cdot 10^{11}$ &
    $1.58 \cdot 10^{07}$ \\

    21 &
    theme-park &
    Theme park vehicles &
    \cite{Forschelen2012} &
    $2.95 \cdot 10^{05}$ &
    $6.69 \cdot 10^{06}$ \\

    22 &
    wafer-scanner-n1 &
    Wafer scanner ($n\!=\!1$) &
    \cite{Sanden2015} &
    $5.30 \cdot 10^{05}$ &
    $5.24 \cdot 10^{04}$ \\

    23 &
    waterway-lock &
    Waterway lock &
    \cite{Reijnen2017} &
    $5.96 \cdot 10^{32}$ &
    $5.87 \cdot 10^{24}$ \\

    \end{tabular}
\end{table*}

\subsection{Benchmark descriptions}

We provide a short description of each of the 23 benchmark models.

\textbf{1)} The \emph{advanced driver assistance system} (ADAS) concerns a safe controller for an adaptive cruise control system for the Toyota Prius Executive, which is controlled via a Human Machine Interface (HMI).
The synthesized controller was tested in multiple ways, including by driving around in real-world traffic~\cite{Korssen2018}.

\textbf{2)} The \emph{automated guided vehicles} (AGVs) serve a manufacturing work cell that produces parts.
The five AGVs travel on fixed circular routes, alternately loading and unloading.
The AGVs move through shared zones, and no two AGVs may be in such a zone at the same time, to avoid collisions~\cite{Wonham2019}.

\textbf{3+4)} The \emph{body comfort system} (BCS) product line is a frequently used benchmark in product line engineering (PLE).
The BCS product line originates from the automotive industry.
There is a \emph{dynamic} version of the model (benchmark 3), which allows the system to be reconfigured during execution, and a \emph{static} version of the model (benchmark 4), which does not allow the system to be reconfigured during execution.
Synthesis will ensure that all safety requirements are satisfied, regardless of the feature configuration~\cite{Beek2016,Thuijsman2023}.

\textbf{5)} The Algera \emph{bridge} over the Hollandse IJssel river connects two cities in the Netherlands.
It is a movable bridge consisting of two vehicle lanes, a lane for cyclists and a lane for pedestrians, as well as various stop signs, traffic barriers, and so on.
A fault-tolerant controller is synthesized to ensure the safe operation of the bridge, even in case of, for instance, faulty sensors or broken actuator connections~\cite{Reijnen2018a}.

\textbf{6+7)} A \emph{cat and mouse tower} (CMT) is a tower that has $n$ levels, with five rooms at each level.
Cats and mice may move between connected rooms, either at the same level or to the level above or below.
The controller must ensure that a cat and a mouse are never in the same room at the same time.
Most moves can be prevented by the supervisor, but not all of them.
In the original version of the model (\emph{version 1}, benchmark 6), $k$ cats start in room 1 at level 1, and $k$ mice in room 5 at level $n$.
In the adapted version of the model (\emph{version 2}, benchmark 7), initially all rooms are empty, cats and mice can enter and leave the tower (cats via room 1 at level 1, and mice via room 5 at level $n$), and at most $k$ cats may be in each room, and similarly at most $k$ mice may be in each room~\cite{Ma2008,Thuijsman2021}.

\textbf{8)} The \emph{cluster tool} is an integrated manufacturing system used for wafer processing.
Wafers enter through the entering load lock and exit through the exiting load lock.
They can be processed in nine chambers.
Four transportation robots transport the wafers.
Three one-slot buffers are used for intermediate storage.
Wafers follow pre-specified routing sequences.
The synthesized controller guarantees continuous wafer processing without blocking situations~\cite{Su2010}.

\textbf{9)} The \emph{dining philosophers} model concerns a simplified version of the traditional dining philosophers problem.
Five philosophers dine at a table, and each has its own plate.
Between the plates are forks.
Each philosopher requires two forks to eat the meal.
No two philosophers can take the same fork at the same time~\cite{Hoare1985}.

\textbf{10)} The \emph{FESTO production line} consists of six connected stations.
Products enter through the distributing station.
They proceed to the handling station, which buffers products, before they go to the testing station.
The testing station measures the height of the products, rejects faulty products, and lets correct products through to the buffering station.
After the buffering station, products are processed by drilling a hole in them.
Finally, products arrive at the storage station.
The supervisor ensures safe, correct, and efficient operation of the product line~\cite{Reijnen2018b}.

\textbf{11)} The \emph{lithography machine initialization} concerns a lithography system with multiple sub-systems, that need to be initialized in the correct order.
Each sub-system is initialized from its unknown starting state to its initialized state.
A sub-system can be initialized via one or more phases, by transitioning from a phase to a previous or next one.
The synthesized supervisor takes into account the dependencies between the system, constraining the initialization order.
If the synthesized supervisor is not empty, a successful initialization sequence exists~\cite{Vos2020,Thuijsman2021}.

\textbf{12+13)} The \emph{patient support system (PSS)} positions a patient in a \emph{Magnetic Resonance Imaging (MRI) scanner}.
It consists primarily of the patient table, on which the patient lies.
The table is moved into a hole in the scanner (the `bore').
A user interface allows medical personnel to operate the patient table.
The table can be positioned manually, but there is also a `light-visor' for automatic positioning.
The synthesized supervisor must ensure that the patient support system operates safely, such that the MRI scanner is not damaged, and that the patient remains safe~\cite{Theunissen2014,Theunissen2015}.
There are two versions of the model, a traditional event-based one (benchmark 12), and one with state-based invariants (benchmark 13).

\textbf{14)} The \emph{multi-agent formation} problem has as goal to control a team of agents.
Agents can only move clockwise around a circular route.
They must assume certain geometric formations, i.e., an equilateral triangle and an alignment curve.
The team leader or remote operator decides the formation to use.
The synthesized supervisor ensures that the desired formations are reached~\cite{Cai2015}.

\textbf{15)} A \emph{production cell} consists of a stock, a feed belt, an elevating rotary table, a robot, a press, a deposit belt, and a crane.
Although they each execute asynchronously and independently, they synchronize for cooperation and safety~\cite{Feng2009}.

\textbf{16--19)} Swarm robotics involves a group of wheeled robots that interact to solve relatively complex tasks cooperatively.
\emph{Swarm robotics aggregation} (benchmark 16) involves a group of robots that gather in a homogeneous environment.
Each robot is equipped with a sensor to detect the presence of other robots in its line of sight.
A robot that does not detect any other robots moves backwards along a circular trajectory.
A robot that does detect another robot turns clockwise.
\emph{Swarm robotics clustering} (benchmark 17) involves that the robots cluster objects that are initially dispersed in the environment.
Robots are equipped with sensors to detect objects and other robots in their direct line of sight, and they move based on their sensor readings.
\emph{Swarm robotics group formation} (benchmark 18) involves leader robots, green follower robots, and blue follower robots.
Each leader should be in equilibrium, with its number of green and blue followers differing by at most one.
Leaders initially stand still at a random location, whereas followers move randomly.
Leaders send offers to followers, which followers may accept, making them stop and relay their leader's messages.
\emph{Swarm robotics segregation} (benchmark 19) also involves leader and follower robots.
Each leader has a type, characterized by a color (red, green, or blue).
Followers that are not in range of a leader, or only in range of a single type of leader, do not move.
Followers that are in range of multiple types of leaders move randomly~\cite{Lopes2016}.

\textbf{20)} \emph{Sudoku} is a number placement game.
Within an (in this case) $4$-by-$4$ grid, each entry needs to be assigned a number.
The game is completed when the top-left, top-right, bottom-left, and bottom-right $2$-by-$2$ grids contain the numbers $1$ through $4$, every row contains these numbers as well, and so does every column.
Synthesis computes all possible solutions.

\textbf{21)} The Multimover is a flexible \emph{theme park vehicle}.
It is used in both theme parks and museums.
The vehicle is battery-operated and follows an electrical wire that is integrated in the floor.
The floor also contains floor codes, that provide information about the track, influence the route and speed of the vehicle, the music it plays, and so on.
An operator controls deployment of the vehicles, while `ride control' coordinates all vehicles, starts, and stops them.
Sensors are used to detect physical objects and prevent collisions.
The synthesized supervisor ensures the safe operation of the vehicles~\cite{Forschelen2012}.

\textbf{22)} The \emph{wafer scanner} is a lithography machine used in the production process of integrated circuits.
Part of the control of the system is the wafer logistics:
the wafer handler takes wafers from the track, performs a number of pre-exposure steps (such as conditioning and alignment), and routes wafers to the wafer stage and back.
At the stage, wafers are put on one of the two chucks, and subsequently they are measured and exposed.
Since immersion lithography is used, a wafer must always be present on the exposure chuck to avoid disruption of the film of water below the lens; dummy wafers are inserted into the system for this.
The supervisor ensures a correct and safe wafer flow through the system.
The model for $n=1$ is a simplified version of the full model, with only one production wafer, besides the two dummy wafers, for at most three wafers in total in the system \cite{Sanden2015}.

\textbf{23)} Lock III is a \emph{waterway lock} in Tilburg, the Netherlands.
The lock features gates, which are watertight doors that are used to seal off the chamber from outside water.
The lock's downstream side features paddles for emptying the chamber, while its upstream side has culverts to fill the chamber.
Vessel traffic is regulated by traffic lights.
In a remote control center, operators monitor the lock via camera images and issue commands via a graphical user interface (GUI) of the Supervisory Control And Data Acquisition (SCADA) system, to for instance open the gates.
The supervisor ensures the safe and correct operation of the lock~\cite{Reijnen2017}.

\subsection{Additional benchmark metrics}

\begin{table*}[b!]
    \caption{Additional metrics about the benchmark models.}
    \label{tbl:benchmark-metrics}
    \centering
    \small
    \setlength{\tabcolsep}{2.45pt}
    \begin{tabular}{ l l rr|rrrrrrrrrr|rrr|rrr|rr|rrrr }
        \textbf{Nr} &
        \textbf{Short name} &
        \multicolumn{2}{c|}{\textbf{Events}} &
        \multicolumn{10}{c|}{\textbf{Automata}} &
        \multicolumn{3}{c|}{\textbf{Initial.}} &
        \multicolumn{3}{c|}{\textbf{Marking}} &
        \multicolumn{2}{c|}{\textbf{Vars.}} &
        \multicolumn{4}{c}{\textbf{Invariants}} \\

        ~ &
        ~ &
        \bm{$\Sigma_c$} & \bm{$\Sigma_u$} &
        \bm{$A_p$} & \bm{$A_r$} &
        \bm{$l_p$} & \bm{$l_r$} &
        \bm{$e_p$} & \bm{$e_r$} &
        \bm{$g_p$} & \bm{$g_r$} & 
        \bm{$a_p$} & \bm{$a_r$} &
        \bm{$i_p$} & \bm{$i_r$} & \bm{$i_c$} &
        \bm{$m_p$} & \bm{$m_r$} & \bm{$m_c$} &
        \bm{$v_n$} & \bm{$v_v$} &
        \bm{$t_{p,s}$} & \bm{$t_{r,s}$} & \bm{$t_{p,e}$} & \bm{$t_{r,e}$} \\

        1 &
adas &
$32$ &
$35$ &
$28$ &
$3$ &
$63$ &
$6$ &
$75$ &
$12$ &
$0$ &
$0$ &
$0$ &
$0$ &
$28$ &
$3$ &
$0$ &
$28$ &
$4$ &
$0$ &
$0$ &
$0$ &
$0$ &
$0$ &
$0$ &
$69$ \\

2 &
agv &
$10$ &
$16$ &
$5$ &
$8$ &
$26$ &
$23$ &
$26$ &
$29$ &
$0$ &
$0$ &
$0$ &
$0$ &
$5$ &
$8$ &
$0$ &
$5$ &
$23$ &
$0$ &
$0$ &
$0$ &
$0$ &
$0$ &
$0$ &
$0$ \\

3 &
bcs-dynamic &
$48$ &
$96$ &
$58$ &
$0$ &
$115$ &
$0$ &
$179$ &
$0$ &
$54$ &
$0$ &
$54$ &
$0$ &
$61$ &
$0$ &
$1$ &
$62$ &
$0$ &
$0$ &
$27$ &
$54$ &
$0$ &
$0$ &
$88$ &
$54$ \\

4 &
bcs-static &
$48$ &
$42$ &
$58$ &
$0$ &
$115$ &
$0$ &
$125$ &
$0$ &
$0$ &
$0$ &
$0$ &
$0$ &
$61$ &
$0$ &
$1$ &
$62$ &
$0$ &
$0$ &
$27$ &
$54$ &
$0$ &
$0$ &
$88$ &
$54$ \\

5 &
bridge &
$65$ &
$149$ &
$80$ &
$0$ &
$181$ &
$0$ &
$229$ &
$0$ &
$0$ &
$0$ &
$27$ &
$0$ &
$80$ &
$0$ &
$0$ &
$88$ &
$0$ &
$0$ &
$21$ &
$42$ &
$0$ &
$0$ &
$136$ &
$189$ \\

6 &
cmt-v1 &
$44$ &
$6$ &
$30$ &
$0$ &
$90$ &
$0$ &
$200$ &
$0$ &
$0$ &
$0$ &
$0$ &
$0$ &
$30$ &
$0$ &
$0$ &
$30$ &
$0$ &
$0$ &
$0$ &
$0$ &
$0$ &
$15$ &
$0$ &
$0$ \\

7 &
cmt-v2 &
$44$ &
$10$ &
$30$ &
$0$ &
$60$ &
$0$ &
$104$ &
$0$ &
$0$ &
$0$ &
$0$ &
$0$ &
$30$ &
$0$ &
$0$ &
$30$ &
$0$ &
$0$ &
$0$ &
$0$ &
$0$ &
$15$ &
$0$ &
$0$ \\

8 &
cluster-tool &
$32$ &
$9$ &
$18$ &
$16$ &
$58$ &
$32$ &
$85$ &
$32$ &
$0$ &
$0$ &
$0$ &
$0$ &
$18$ &
$16$ &
$0$ &
$18$ &
$16$ &
$0$ &
$0$ &
$0$ &
$0$ &
$0$ &
$0$ &
$0$ \\

9 &
dining-phils &
$15$ &
$0$ &
$10$ &
$0$ &
$30$ &
$0$ &
$45$ &
$0$ &
$0$ &
$0$ &
$0$ &
$0$ &
$10$ &
$0$ &
$0$ &
$10$ &
$0$ &
$0$ &
$0$ &
$0$ &
$0$ &
$0$ &
$0$ &
$0$ \\

10 &
festo &
$64$ &
$122$ &
$91$ &
$3$ &
$186$ &
$16$ &
$190$ &
$18$ &
$0$ &
$0$ &
$0$ &
$0$ &
$91$ &
$3$ &
$0$ &
$95$ &
$3$ &
$0$ &
$0$ &
$0$ &
$0$ &
$0$ &
$25$ &
$200$ \\

11 &
litho-init &
$68$ &
$98$ &
$16$ &
$0$ &
$212$ &
$0$ &
$324$ &
$0$ &
$0$ &
$0$ &
$0$ &
$0$ &
$46$ &
$0$ &
$0$ &
$16$ &
$0$ &
$0$ &
$0$ &
$0$ &
$0$ &
$50$ &
$0$ &
$0$ \\

12 &
mri-pss-event &
$20$ &
$19$ &
$17$ &
$17$ &
$47$ &
$46$ &
$105$ &
$112$ &
$0$ &
$0$ &
$0$ &
$0$ &
$17$ &
$17$ &
$0$ &
$41$ &
$44$ &
$0$ &
$0$ &
$0$ &
$0$ &
$0$ &
$0$ &
$0$ \\

13 &
mri-pss-state &
$13$ &
$19$ &
$12$ &
$3$ &
$32$ &
$7$ &
$47$ &
$15$ &
$0$ &
$0$ &
$0$ &
$0$ &
$18$ &
$3$ &
$0$ &
$22$ &
$5$ &
$0$ &
$0$ &
$0$ &
$3$ &
$2$ &
$8$ &
$12$ \\

14 &
multi-agent-form &
$30$ &
$2$ &
$4$ &
$13$ &
$31$ &
$26$ &
$32$ &
$29$ &
$2$ &
$0$ &
$0$ &
$0$ &
$4$ &
$13$ &
$0$ &
$28$ &
$25$ &
$1$ &
$0$ &
$0$ &
$0$ &
$0$ &
$0$ &
$0$ \\

15 &
prod-cell &
$27$ &
$48$ &
$11$ &
$19$ &
$81$ &
$55$ &
$87$ &
$61$ &
$0$ &
$0$ &
$0$ &
$0$ &
$11$ &
$19$ &
$0$ &
$11$ &
$19$ &
$0$ &
$0$ &
$0$ &
$0$ &
$0$ &
$0$ &
$0$ \\

16 &
robo-swarm-aggr &
$2$ &
$2$ &
$2$ &
$4$ &
$2$ &
$8$ &
$2$ &
$16$ &
$0$ &
$0$ &
$0$ &
$0$ &
$2$ &
$4$ &
$0$ &
$2$ &
$8$ &
$0$ &
$0$ &
$0$ &
$0$ &
$0$ &
$0$ &
$0$ \\

17 &
robo-swarm-clus &
$3$ &
$3$ &
$2$ &
$6$ &
$2$ &
$12$ &
$2$ &
$24$ &
$0$ &
$0$ &
$0$ &
$0$ &
$2$ &
$6$ &
$0$ &
$2$ &
$12$ &
$0$ &
$0$ &
$0$ &
$0$ &
$0$ &
$0$ &
$0$ \\

18 &
robo-swarm-form &
$17$ &
$11$ &
$6$ &
$6$ &
$14$ &
$22$ &
$15$ &
$38$ &
$0$ &
$0$ &
$0$ &
$0$ &
$6$ &
$6$ &
$0$ &
$10$ &
$22$ &
$0$ &
$0$ &
$0$ &
$0$ &
$0$ &
$0$ &
$0$ \\

19 &
robo-swarm-segr &
$8$ &
$8$ &
$6$ &
$3$ &
$13$ &
$8$ &
$14$ &
$17$ &
$0$ &
$0$ &
$0$ &
$0$ &
$6$ &
$3$ &
$0$ &
$6$ &
$6$ &
$0$ &
$0$ &
$0$ &
$0$ &
$0$ &
$0$ &
$0$ \\

20 &
sudoku &
$64$ &
$0$ &
$16$ &
$0$ &
$16$ &
$0$ &
$64$ &
$0$ &
$64$ &
$0$ &
$64$ &
$0$ &
$16$ &
$0$ &
$0$ &
$16$ &
$0$ &
$48$ &
$16$ &
$80$ &
$0$ &
$0$ &
$0$ &
$0$ \\

21 &
theme-park &
$22$ &
$25$ &
$17$ &
$5$ &
$36$ &
$19$ &
$47$ &
$77$ &
$0$ &
$0$ &
$0$ &
$0$ &
$17$ &
$5$ &
$0$ &
$17$ &
$5$ &
$0$ &
$0$ &
$0$ &
$0$ &
$0$ &
$0$ &
$44$ \\

22 &
wafer-scanner-n1 &
$75$ &
$79$ &
$26$ &
$10$ &
$192$ &
$26$ &
$287$ &
$27$ &
$7$ &
$0$ &
$4$ &
$0$ &
$26$ &
$10$ &
$0$ &
$57$ &
$10$ &
$0$ &
$1$ &
$4$ &
$0$ &
$0$ &
$0$ &
$31$ \\

23 &
waterway-lock &
$94$ &
$170$ &
$71$ &
$0$ &
$230$ &
$0$ &
$382$ &
$0$ &
$0$ &
$0$ &
$0$ &
$0$ &
$71$ &
$0$ &
$0$ &
$73$ &
$0$ &
$0$ &
$0$ &
$0$ &
$0$ &
$0$ &
$116$ &
$198$ \\

    \end{tabular}
\end{table*}

Table~\ref{tbl:benchmark-metrics} shows the following additional metrics for each of the benchmark models:

\begin{description}[leftmargin=!, labelwidth=0.5cm, noitemsep]
    \setlength\itemsep{0em}
    \item[$\Sigma_c$] The number of controllable events.
    \item[$\Sigma_u$] The number of uncontrollable events.
    \item[$A_p$] The number of plant automata.
    \item[$A_r$] The number of requirement automata.
    \item[$l_p$] The number of plant locations, summed over the automata of $A_p$.
    \item[$l_r$] The number of requirement locations, summed over the automata of $A_r$.
    \item[$e_p$] The number of plant edges, summed over the locations of $l_p$. If a single edge is labeled with multiple events, as shorter notation for writing multiple edges, then the edge is counted as many times as there are events on the edge.
    \item[$e_r$] The number of requirement edges, summed over the locations of $l_r$. Similarly counted as $e_p$.
    \item[$g_p$] The number of plant guards, i.e., the number of edges in $e_p$ that have a non-\emph{true} guard.
    \item[$g_r$] The number of requirement guards, i.e., the number of edges in $e_r$ that have a non-\emph{true} guard.
    \item[$a_p$] The number of plant assignments, i.e., the number of assignments on edges in $e_p$. A single edge may have multiple assignments for the same variable, if conditional updates are used, and these are all counted separately.
    \item[$a_r$] The number of requirement assignments, i.e., the number of assignments on edges in $e_r$. Similarly counted as $a_p$.
    \item[$i_p$] The number of initial plant locations, i.e., the locations in $l_p$ that could be initial locations. In CIF, a location can have initialization predicates that indicate under which condition the location is an initial location, for instance in terms of the values of variables or the initial locations of other automata. Each potential initial location is counted once.
    \item[$i_r$] The number of initial requirement locations. Similarly counted as $i_p$.
    \item[$i_c$] The number of initialization predicates in components.
    \item[$m_p$] The number of marked plant locations. Similarly counted as $i_p$.
    \item[$m_r$] The number of marked requirement locations. Similarly counted as $i_r$.
    \item[$m_c$] The number of marker predicates in components. Similarly counted as $i_c$.
    \item[$v_n$] The number of CIF variables.
    \item[$v_v$] The number of values for the CIF variables combined. For each variable, the number of potential values in their CIF ranges are counted, and summed up.
    \item[$t_{p,s}$] The number of plant state invariants.
    \item[$t_{r,s}$] The number of requirement state invariants.
    \item[$t_{p,e}$] The number of plant state/event exclusion invariants.
    \item[$t_{r,e}$] The number of requirement state/event exclusion invariants.
\end{description}

\section{Experimental evaluation of symbolic synthesis performance improvements}
\label{sec:experiments}

To show the cumulative effect of the performance improvements described in Section~\ref{sec:improvements}, and determine the practical improvements to synthesis performance that practitioners may expect, we perform experiments.
Besides the relative delta in performance due to the improvements, we also evaluate the current synthesis performance (the absolute time that synthesis takes), to show researchers and practitioners what they may expect in terms of synthesis performance.

The artifact that accompanies this paper includes the scripts to reproduce the data of all tables and figures from this section of the paper~\cite{ArtifactOfThisPaper}.

\subsection{Experimental setup}

The synthesis performance improvements described in Section~\ref{sec:improvements} cover the changes made to CIF's symbolic synthesis tool between ESCET releases v0.8 and v4.0.
We therefore compare the synthesis performance of these two versions of the synthesis tool.
This results in a comparison of the cumulative impact of the improvements, which are all enabled by default in ESCET release v4.0, and are thus available out-of-the-box.
No other performance improvements are present in v4.0 compared to v0.8, except for ones that need to be explicitly enabled through options, which we do not consider.

We perform synthesis for each benchmark model from Section~\ref{sec:benchmarks}, for both versions of the toolkit.
We thereby measure the execution time and memory use.
Since synthesis performance is predominantly determined by the creation of BDDs and operations on BDDs, we use BDD-related time and memory metrics~\cite{Thuijsman2019}.
Such metrics are platform independent.
This allows the experiments to be repeated on different hardware, regardless of the use of a simple laptop or a supercomputer.
It also allows to compare the performance described in this paper with future improvements, even if they will be made years into the future.

To measure time, we count the number of BDD operations performed.
Recall that BDD operations are implemented as recursive functions.
We count each recursive invocation of such functions, but do not count trivial computations, namely those that can be obtained from the cache of the BDD library and those that require no further recursive processing.

To measure memory, we count the maximum number of BDD nodes in use, during synthesis.
We do not count nodes that have been allocated but are no longer in use, waiting to be garbage collected.
This gives an accurate measure of the memory usage of synthesis~\cite{Thuijsman2019}.

\subsection{Results}

\begin{table*}[t!]
    \caption{Results from the experiments showing the detailed effect of the performance improvements.}
    \label{tbl:experiment-results}
    \centering
    \small
    \setlength{\tabcolsep}{7pt}
    \begin{tabular}{ l l c c c c c c c }

        \textbf{Nr} &
        \textbf{Short name} &
        \multicolumn{3}{c}{\textbf{Time [number of operations]}} &
        &
        \multicolumn{3}{c}{\textbf{Memory [maximum number of nodes]}} \\
        \cline{3-5} \cline{7-9}

        ~ &
        ~ &
        \textbf{v0.8} & \textbf{v4.0} & \textbf{Reduction} &
        &
        \textbf{v0.8} & \textbf{v4.0} & \textbf{Reduction} \\

1 &
adas &
$2.2 \cdot 10^{05}$ &
$3.0 \cdot 10^{04}$ &
$\phantom{0}\phantom{0}\phantom{0}7.4$ &
&
$7.0 \cdot 10^{03}$ &
$1.5 \cdot 10^{03}$ &
$\phantom{0}\phantom{0}\phantom{0}4.5$ \\

2 &
agv &
$1.2 \cdot 10^{06}$ &
$2.3 \cdot 10^{05}$ &
$\phantom{0}\phantom{0}\phantom{0}5.4$ &
&
$5.4 \cdot 10^{03}$ &
$2.5 \cdot 10^{03}$ &
$\phantom{0}\phantom{0}\phantom{0}2.2$ \\

3 &
bcs-dynamic &
$2.4 \cdot 10^{07}$ &
$1.1 \cdot 10^{05}$ &
$\phantom{0}223.9$ &
&
$2.7 \cdot 10^{05}$ &
$2.9 \cdot 10^{03}$ &
$\phantom{0}\phantom{0}91.2$ \\

4 &
bcs-static &
$6.1 \cdot 10^{08}$ &
$1.4 \cdot 10^{05}$ &
$4293.6$ &
&
$3.1 \cdot 10^{06}$ &
$5.9 \cdot 10^{03}$ &
$\phantom{0}518.3$ \\

5 &
bridge &
$1.2 \cdot 10^{07}$ &
$1.6 \cdot 10^{06}$ &
$\phantom{0}\phantom{0}\phantom{0}7.5$ &
&
$8.4 \cdot 10^{04}$ &
$2.4 \cdot 10^{04}$ &
$\phantom{0}\phantom{0}\phantom{0}3.5$ \\

6 &
cmt-v1 &
$3.1 \cdot 10^{06}$ &
$4.3 \cdot 10^{05}$ &
$\phantom{0}\phantom{0}\phantom{0}7.3$ &
&
$2.2 \cdot 10^{04}$ &
$1.6 \cdot 10^{04}$ &
$\phantom{0}\phantom{0}\phantom{0}1.4$ \\

7 &
cmt-v2 &
$8.0 \cdot 10^{06}$ &
$3.3 \cdot 10^{06}$ &
$\phantom{0}\phantom{0}\phantom{0}2.4$ &
&
$2.5 \cdot 10^{04}$ &
$1.2 \cdot 10^{04}$ &
$\phantom{0}\phantom{0}\phantom{0}2.1$ \\

8 &
cluster-tool &
$3.3 \cdot 10^{06}$ &
$4.7 \cdot 10^{05}$ &
$\phantom{0}\phantom{0}\phantom{0}7.0$ &
&
$1.2 \cdot 10^{04}$ &
$7.2 \cdot 10^{03}$ &
$\phantom{0}\phantom{0}\phantom{0}1.7$ \\

9 &
dining-phils &
$2.2 \cdot 10^{04}$ &
$6.5 \cdot 10^{03}$ &
$\phantom{0}\phantom{0}\phantom{0}3.4$ &
&
$1.0 \cdot 10^{03}$ &
$4.8 \cdot 10^{02}$ &
$\phantom{0}\phantom{0}\phantom{0}2.2$ \\

10 &
festo &
$8.1 \cdot 10^{08}$ &
$4.8 \cdot 10^{05}$ &
$1672.9$ &
&
$3.6 \cdot 10^{06}$ &
$6.1 \cdot 10^{03}$ &
$\phantom{0}598.5$ \\

11 &
litho-init &
$2.3 \cdot 10^{07}$ &
$5.7 \cdot 10^{06}$ &
$\phantom{0}\phantom{0}\phantom{0}4.0$ &
&
$4.0 \cdot 10^{04}$ &
$3.7 \cdot 10^{04}$ &
$\phantom{0}\phantom{0}\phantom{0}1.1$ \\

12 &
mri-pss-event &
$5.6 \cdot 10^{07}$ &
$1.4 \cdot 10^{07}$ &
$\phantom{0}\phantom{0}\phantom{0}4.1$ &
&
$5.6 \cdot 10^{05}$ &
$7.5 \cdot 10^{04}$ &
$\phantom{0}\phantom{0}\phantom{0}7.5$ \\

13 &
mri-pss-state &
$7.5 \cdot 10^{04}$ &
$2.1 \cdot 10^{04}$ &
$\phantom{0}\phantom{0}\phantom{0}3.6$ &
&
$3.1 \cdot 10^{03}$ &
$1.0 \cdot 10^{03}$ &
$\phantom{0}\phantom{0}\phantom{0}3.0$ \\

14 &
multi-agent-form &
$1.7 \cdot 10^{06}$ &
$4.0 \cdot 10^{05}$ &
$\phantom{0}\phantom{0}\phantom{0}4.3$ &
&
$2.6 \cdot 10^{04}$ &
$2.0 \cdot 10^{04}$ &
$\phantom{0}\phantom{0}\phantom{0}1.3$ \\

15 &
prod-cell &
$9.9 \cdot 10^{07}$ &
$1.3 \cdot 10^{07}$ &
$\phantom{0}\phantom{0}\phantom{0}7.4$ &
&
$3.0 \cdot 10^{04}$ &
$1.7 \cdot 10^{04}$ &
$\phantom{0}\phantom{0}\phantom{0}1.8$ \\

16 &
robo-swarm-aggr &
$5.5 \cdot 10^{02}$ &
$3.5 \cdot 10^{02}$ &
$\phantom{0}\phantom{0}\phantom{0}1.5$ &
&
$1.7 \cdot 10^{02}$ &
$6.8 \cdot 10^{01}$ &
$\phantom{0}\phantom{0}\phantom{0}2.5$ \\

17 &
robo-swarm-clus &
$2.4 \cdot 10^{03}$ &
$1.2 \cdot 10^{03}$ &
$\phantom{0}\phantom{0}\phantom{0}2.0$ &
&
$5.7 \cdot 10^{02}$ &
$1.7 \cdot 10^{02}$ &
$\phantom{0}\phantom{0}\phantom{0}3.4$ \\

18 &
robo-swarm-form &
$3.8 \cdot 10^{04}$ &
$6.8 \cdot 10^{03}$ &
$\phantom{0}\phantom{0}\phantom{0}5.6$ &
&
$4.3 \cdot 10^{03}$ &
$6.5 \cdot 10^{02}$ &
$\phantom{0}\phantom{0}\phantom{0}6.6$ \\

19 &
robo-swarm-segr &
$1.3 \cdot 10^{04}$ &
$2.6 \cdot 10^{03}$ &
$\phantom{0}\phantom{0}\phantom{0}5.1$ &
&
$8.6 \cdot 10^{02}$ &
$2.5 \cdot 10^{02}$ &
$\phantom{0}\phantom{0}\phantom{0}3.5$ \\

20 &
sudoku &
$1.6 \cdot 10^{07}$ &
$6.7 \cdot 10^{06}$ &
$\phantom{0}\phantom{0}\phantom{0}2.4$ &
&
$6.1 \cdot 10^{05}$ &
$5.2 \cdot 10^{05}$ &
$\phantom{0}\phantom{0}\phantom{0}1.2$ \\

21 &
theme-park &
$4.1 \cdot 10^{05}$ &
$2.6 \cdot 10^{04}$ &
$\phantom{0}\phantom{0}15.3$ &
&
$1.3 \cdot 10^{04}$ &
$1.2 \cdot 10^{03}$ &
$\phantom{0}\phantom{0}11.6$ \\

22 &
wafer-scanner-n1 &
$1.2 \cdot 10^{09}$ &
$5.7 \cdot 10^{08}$ &
$\phantom{0}\phantom{0}\phantom{0}2.1$ &
&
$3.6 \cdot 10^{05}$ &
$7.9 \cdot 10^{05}$ &
$\phantom{0}\phantom{0}\phantom{0}0.5$ \\

23 &
waterway-lock &
$5.5 \cdot 10^{07}$ &
$6.8 \cdot 10^{06}$ &
$\phantom{0}\phantom{0}\phantom{0}8.1$ &
&
$9.7 \cdot 10^{04}$ &
$1.4 \cdot 10^{04}$ &
$\phantom{0}\phantom{0}\phantom{0}7.1$ \\

\multicolumn{2}{l}{\textbf{Average (per column)}} &
$1.3 \cdot 10^{08}$ &
$2.7 \cdot 10^{07}$ &
$\pmb{\phantom{0}273.7}$ &
&
$3.9 \cdot 10^{05}$ &
$6.7 \cdot 10^{04}$ &
$\pmb{\phantom{0}\phantom{0}55.5}$ \\

\multicolumn{2}{l}{\textbf{Median (per column)}} &
$3.3 \cdot 10^{06}$ &
$4.0 \cdot 10^{05}$ &
$\pmb{\phantom{0}\phantom{0}\phantom{0}5.4}$ &
&
$2.5 \cdot 10^{04}$ &
$6.1 \cdot 10^{03}$ &
$\pmb{\phantom{0}\phantom{0}\phantom{0}3.0}$ \\

    \end{tabular}
\end{table*}

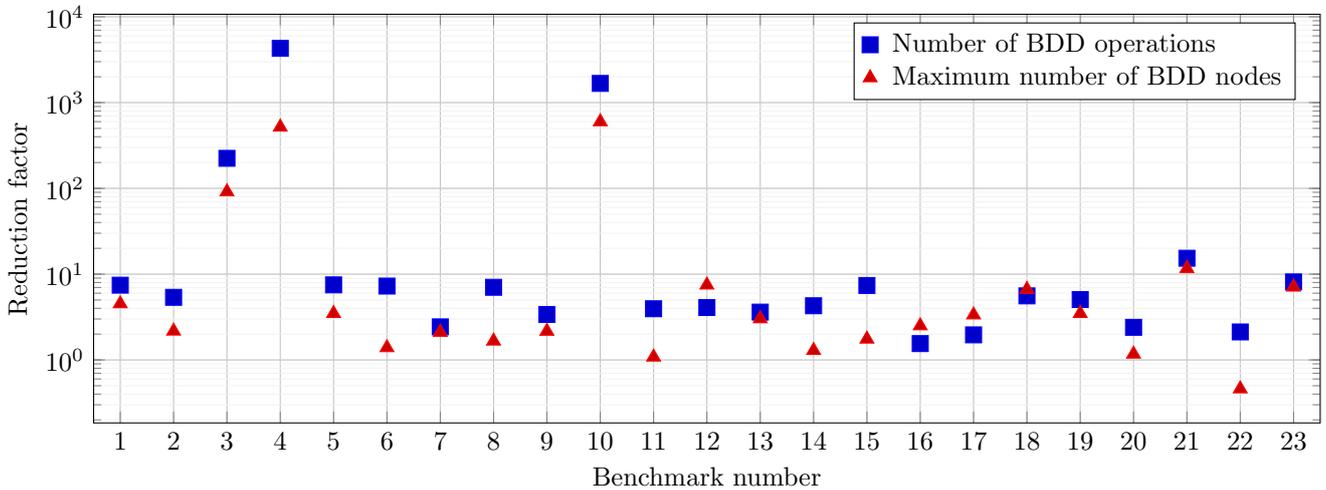
\begin{figure*}[t!]
    \centering
    \begin{tikzpicture}[scale=1.0]
        \begin{axis}[
            width=.99\linewidth, height=7cm,
            xmin=0.5, xmax=23.5,
            xtick={1, 2, 3, 4, 5, 6, 7, 8, 9, 10, 11, 12, 13, 14, 15, 16, 17, 18, 19, 20, 21, 22, 23},
            xlabel = {Benchmark number},
            ylabel = {Reduction factor},
            ymode=log,
            grid=both,
            grid style={line width=.1pt, draw=gray!10},
            major grid style={line width=.2pt,draw=gray!50},
            legend cell align={left},
            legend entries={
                {~Number of BDD operations},
                {~Maximum number of BDD nodes},
            },
            only marks,
            mark size=3pt,
        ]
        \addplot+[mark=square*]
        table [x=nr, y=ops]{images/experiment-results.dat};
        \addplot+[mark=triangle*]
        table [x=nr, y=nodes]{images/experiment-results.dat};
        \end{axis}
    \end{tikzpicture}
    \caption{Results from the experiments showing a graphical overview of the effect of the performance improvements.}
    \label{fig:experiment-results}
\end{figure*}

The results of the experiments are shown in Table~\ref{tbl:experiment-results}.
For each benchmark, the time and memory are indicated, for each ESCET version.
Additionally, the effort reduction factor is indicated.
For instance, a factor of $2.0$ means half the number of operations or half the number of nodes, while a factor of $0.5$ means twice the number of operations or twice the number of nodes.
For $o_1$ the number of operations for ESCET v0.8 and $o_2$ the number of operations for ESCET v4.0, the reduction factor is computed as $\frac{o_1}{o_2}$.
Similarly, for $n_1$ the maximum number of nodes for ESCET v0.8 and $n_2$ the maximum number of nodes for ESCET v4.0, the reduction factor is computed as $\frac{n_1}{n_2}$.
A more graphical overview of the same results is shown in Figure~\ref{fig:experiment-results}, where the reduction factors per benchmark are plotted using a scatter plot, with a logarithmic vertical axis.

The time (in number of BDD operations) has reduced in v4.0 compared to v0.8, for all benchmarks.
On average, the time is reduced by a factor of $274$, more than two orders of magnitude.
The median is a reduction factor of only $5.4$, since a few benchmarks with much higher reduction factors increase the average quite a bit.
The maximum reduction is by a factor of $4{,}294$, but for most models the reduction factor is less than $10$.
The result thus greatly varies between benchmarks, but overall it takes significantly less time to perform synthesis.

The memory usage (in maximum number of BDD nodes) is on average reduced by a factor of $56$, considerably more than an order of magnitude.
The median is a reduction factor of only $3.0$, since here too a few benchmarks with much higher reduction factors increase the average quite a bit.
The maximum reduction is by a factor of $598$, but for most models the reduction factor is less than $8$.
One benchmark has a reduction factor less than one, and thus synthesis actually uses more memory: for the wafer scanner, the maximum number of BDD nodes has approximately doubled.
Here too, the result thus greatly varies between benchmarks, but overall it takes significantly less memory to perform synthesis.

That the results differ between benchmarks is not surprising.
One reason is that the effort that synthesis needs to spend simply differs for the problem being solved.
Another reason is that various improvements are based on heuristics, such as the variable ordering algorithms.
The heuristics are tuned to work well for most models, but there can always be models where performance is adversely affected.

\begin{table}[t!]
    \caption{Impact of further optimizing synthesis settings on the performance for the wafer scanner benchmark.}
    \label{tbl:experiment-optimized}
    \centering
    \small
    \setlength{\tabcolsep}{15pt}
    \begin{tabular}{ l c c }

        \textbf{Version} &
        \textbf{Operations} &
        \textbf{Nodes} \\

        ESCET v0.8 &
        $1.2 \cdot 10^{09}$ & %1.20
        $3.6 \cdot 10^{05}$ \\ %3.63

        ESCET v4.0 (default config.) &
        $5.7 \cdot 10^{08}$ & %5.67
        $7.9 \cdot 10^{05}$ \\ %7.87

        ESCET v4.0 (optimized config.) &
        $3.4 \cdot 10^{07}$ & %34332207
        $8.4 \cdot 10^{04}$ \\ %83899

    \end{tabular}
\end{table}

CIF's synthesis algorithm can be customized through many settings.
This includes configuration of the various heuristics.
By experimenting with these settings, the synthesis performance may be further improved for specific models.
For instance, for the wafer scanner benchmark, by further tweaking the variable ordering algorithms, enabling forward reachability, computing the reachable states first (before non-blocking and controllable states), and changing the order in which the edges are applied during reachability computations, we can reduce both the number of BDD operations and the maximum number of used BDD nodes, as shown in Table~\ref{tbl:experiment-optimized}.
In this case, the number of BDD operations and the maximum number of BDD nodes are both reduced by about an order of magnitude.

\subsection{Current synthesis performance}
\label{sec:current-synthesis-performance}

So far, in this section, we showed the relative improvements to the synthesis performance, as a delta in performance between ESCET v0.8 and v4.0.
To conclude this section, we now show the current synthesis performance in v4.0.

Unlike before, we do not use the BDD-related performance metrics.
Instead, we measure the absolute amount of time in seconds it takes to synthesize the benchmarks with ESCET v4.0, using the default configuration of the synthesis tool.
This gives an indication on the current state of the efficiency of synthesis for practical application.
For each benchmark, synthesis is performed ten times, and the average synthesis time in seconds is computed.
Each benchmark is synthesized once before the actual measurements, to warm up the Java Virtual Machine (JVM), resulting in more stable measurements.
The experiments are performed on a laptop with a 3 GHz Intel Core i7-1185G7 processor.

Table~\ref{tbl:experiment-times} shows the results.
The wafer scanner benchmark takes by far the longest: a bit more than one and a half minute.
The other benchmarks can each be synthesized within just a few seconds, and most of them even within a second.

\begin{table}[b!]
    \caption{The absolute time in seconds it takes to synthesize the benchmarks using ESCET v4.0, with the default configuration.}
    \label{tbl:experiment-times}
    \centering
    \small
    \setlength{\tabcolsep}{10pt}
    \begin{tabular}{ l l r | l l r }

        \textbf{Nr} &
        \textbf{Short name} &
        \textbf{Time} &
        \textbf{Nr} &
        \textbf{Short name} &
        \textbf{Time} \\

        1 & adas & $0.04$ &
13 & mri-pss-state & $0.02$ \\
2 & agv & $0.05$ &
14 & multi-agent-form & $0.08$ \\
3 & bcs-dynamic & $0.08$ &
15 & prod-cell & $0.98$ \\
4 & bcs-static & $0.08$ &
16 & robo-swarm-aggr & $0.01$ \\
5 & bridge & $0.30$ &
17 & robo-swarm-clus & $0.02$ \\
6 & cmt-v1 & $0.09$ &
18 & robo-swarm-form & $0.02$ \\
7 & cmt-v2 & $0.34$ &
19 & robo-swarm-segr & $0.01$ \\
8 & cluster-tool & $0.07$ &
20 & sudoku & $4.40$ \\
9 & dining-phils & $0.02$ &
21 & theme-park & $0.04$ \\
10 & festo & $0.14$ &
22 & wafer-scanner-n1 & $97.03$ \\
11 & litho-init & $0.53$ &
23 & waterway-lock & $0.73$ \\
12 & mri-pss-event & $1.93$ &
~ & ~ & ~ \\

    \end{tabular}
\end{table}

\subsection{Threats to validity}

There is a risk of selection bias, since the benchmarks that we use come for a large part from our own academic community (of which we are a part) and from the companies that we work with.
Given the lack of other suitable benchmark sets in the literature, this can't be avoided.

More specifically, there is a risk of sampling bias, since we can't guarantee that the benchmarks we've used are representative of the real-world models used in industry.
We have partially mitigated this risk by including various industrial models as benchmarks.

Furthermore, various algorithms involve heuristics that were tuned, to a large extent using the same benchmarks.
To mitigate this risk, additional benchmarks were added later on, such as the MRI and wafer scanner ones.

Finally, there is some risk of attrition bias, as for some benchmarks, such as CMT and wafer scanner, lower parameter values were used to ensure that the benchmarks could be synthesized within reasonable time in older ESCET releases.
As most benchmarks do not have parameters, this significantly reduces this risk.
In ESCET v4.0, most benchmarks can be easily synthesized in a short time.
It may thus be time to consider adding more challenging benchmarks.

\section{Non-monolithic synthesis}
\label{sec:non-monolithic}

Despite the performance improvements over the years, and the ones described in Section~\ref{sec:improvements}, scalability remains a concern for the practical application of supervisory controller synthesis.
For instance, the wafer scanner benchmark model as discussed in Section~\ref{sec:benchmarks} is used in this paper with parameter $n\!=\!1$.
To obtain the complete system behavior, we would need to use $n\!=\!7$.
Then seven production wafers and two dummy wafers, for a total of nine wafers, can be in the system at a time.
However, synthesis time and memory usage increase significantly when we increase $n$.
As mentioned in Section~\ref{sec:current-synthesis-performance}, the model for $n\!=\!1$ can be synthesized on a laptop in under two minutes using the default settings.
And with tuning to optimize the settings, we can synthesize it in about four seconds.
We were able to also synthesize a supervisor for $n\!=\!2$, using the same tuned settings as for $n\!=\!1$, in about ten minutes.
For $n\!=\!3$, again using the same tuned settings, we were able to synthesize a supervisor in a bit more than eight hours.
We have so far not been able to synthesize a supervisor for $n\!=\!4$, even through tuning with different settings, as either synthesis runs out of memory, or is still running after 100 hours.
Nonetheless, this is significantly better than what van der Sanden et al.~originally reported in 2015~\cite{Sanden2015}.
They were able to synthesize a supervisor for $n\!=\!1$, but they could only go to $n\!=\!2$ by significantly reducing the model's complexity.
The artifact that accompanies this paper includes the models and scripts we used to synthesize the wafer scanner benchmark for $1\leq n\leq 3$~\cite{ArtifactOfThisPaper}.

Although we believe further performance improvements to be possible, the question is whether they would be sufficient to resolve the seemingly exponentially increasing synthesis times for the wafer scanner benchmark.
So far, we have only discussed monolithic supervisory controller synthesis, where based on the plants and the requirements a single supervisory controller is synthesized.
An alternative is to use non-monolithic synthesis approaches, that synthesize multiple supervisors for parts of the system, which together ensure the correct and safe operation of the system.
Various techniques have been proposed over the years, such as modular~\cite{Wonham1988,Queiroz2000,Malik2016}, decentralized~\cite{Lin1988,Rudie1991,Lee2022}, hierarchical~\cite{Zhong1990,Wong1996} and compositional~\cite{Flordal2007} synthesis, as well as combinations of them~\cite{Schmidt2004,Hill2006}.

Relatively recent is multilevel synthesis~\cite{Komenda2016}, where plant components and requirements are automatically grouped together into a hierarchical structure, the multilevel system, and a separate supervisor is synthesized for each group of this multilevel system.
This allows us to distribute the control problem over multiple cooperating supervisors, which together are significantly smaller than one monolithic supervisor.
By encoding the relations between plant components and requirements in a design structure matrix (DSM), and algorithmically reordering its rows and columns to place tightly coupled plant components side by side~\cite{Wilschut2017}, a suitable multilevel structure can be obtained (semi-\nobreak)\allowbreak{}automatically.
Compared to monolithic synthesis, this can substantially reduce synthesis effort for certain systems~\cite{Goorden2020}, allowing synthesis for much larger variants of such systems.

\begin{table*}[b!]
    \caption{Results of multilevel synthesis compared to monolithic synthesis, for supported benchmarks.}
    \label{tbl:experiment-multilevel}
    \centering
    \small
    \setlength{\tabcolsep}{5pt}
    \begin{tabular}{ ll r | ccc | ccc }

    \textbf{Nr} &
    \textbf{Short name} &
    \textbf{Nr. of} &
    \multicolumn{3}{c|}{\textbf{Number of operations}} &
    \multicolumn{3}{c}{\textbf{Maximum number of nodes}} \\

    ~ &
    ~ &
    \textbf{groups} &
    \textbf{Monolithic} &
    \textbf{Multilevel} &
    \textbf{Reduction} &
    \textbf{Monolithic} &
    \textbf{Multilevel} &
    \textbf{Reduction} \\

    1 &
adas &
34 &
$3.0 \cdot 10^{04}$ &
$2.1 \cdot 10^{04}$ &
$\phantom{0}1.4$ &
$1.5 \cdot 10^{03}$ &
$1.1 \cdot 10^{03}$ &
$\phantom{0}1.4$ \\

2 &
agv &
8 &
$2.3 \cdot 10^{05}$ &
$3.3 \cdot 10^{04}$ &
$\phantom{0}6.9$ &
$2.5 \cdot 10^{03}$ &
$6.6 \cdot 10^{02}$ &
$\phantom{0}3.8$ \\

8 &
cluster-tool &
1 &
$4.7 \cdot 10^{05}$ &
$4.7 \cdot 10^{05}$ &
$\phantom{0}1.0$ &
$7.2 \cdot 10^{03}$ &
$7.2 \cdot 10^{03}$ &
$\phantom{0}1.0$ \\

12 &
mri-pss-event &
7 &
$1.4 \cdot 10^{07}$ &
$3.3 \cdot 10^{05}$ &
$42.2$ &
$7.5 \cdot 10^{04}$ &
$6.2 \cdot 10^{03}$ &
$12.1$ \\

15 &
prod-cell &
11 &
$1.3 \cdot 10^{07}$ &
$1.3 \cdot 10^{07}$ &
$\phantom{0}1.0$ &
$1.7 \cdot 10^{04}$ &
$1.7 \cdot 10^{04}$ &
$\phantom{0}1.0$ \\

16 &
robo-swarm-aggr &
3 &
$3.5 \cdot 10^{02}$ &
$1.8 \cdot 10^{02}$ &
$\phantom{0}2.0$ &
$6.8 \cdot 10^{01}$ &
$2.6 \cdot 10^{01}$ &
$\phantom{0}2.6$ \\

17 &
robo-swarm-clus &
3 &
$1.2 \cdot 10^{03}$ &
$5.9 \cdot 10^{02}$ &
$\phantom{0}2.0$ &
$1.7 \cdot 10^{02}$ &
$5.3 \cdot 10^{01}$ &
$\phantom{0}3.2$ \\

18 &
robo-swarm-form &
9 &
$6.8 \cdot 10^{03}$ &
$5.0 \cdot 10^{03}$ &
$\phantom{0}1.3$ &
$6.5 \cdot 10^{02}$ &
$4.4 \cdot 10^{02}$ &
$\phantom{0}1.5$ \\

19 &
robo-swarm-segr &
9 &
$2.6 \cdot 10^{03}$ &
$1.8 \cdot 10^{03}$ &
$\phantom{0}1.4$ &
$2.5 \cdot 10^{02}$ &
$1.4 \cdot 10^{02}$ &
$\phantom{0}1.8$ \\

21 &
theme-park &
22 &
$2.6 \cdot 10^{04}$ &
$8.2 \cdot 10^{03}$ &
$\phantom{0}3.2$ &
$1.2 \cdot 10^{03}$ &
$5.6 \cdot 10^{02}$ &
$\phantom{0}2.1$ \\

22 &
wafer-scanner-n1 &
1 &
$5.7 \cdot 10^{08}$ &
$5.7 \cdot 10^{08}$ &
$\phantom{0}1.0$ &
$7.9 \cdot 10^{05}$ &
$7.9 \cdot 10^{05}$ &
$\phantom{0}1.0$ \\

    \end{tabular}
\end{table*}

To show how much multilevel synthesis can help in practice, we again use the set of benchmark models from Section~\ref{sec:benchmarks}.
We use the under-development multilevel synthesis tool from ESCET v4.0 to generate specifications for the various groups of the multilevel system.
We then use ESCET v4.0 to synthesize a supervisor for each of the groups.
The artifact that accompanies this paper includes the scripts we used for this experiment~\cite{ArtifactOfThisPaper}.
Out of 23 benchmarks, currently only 11 are supported by CIF's multilevel synthesis tool.
The other 12 benchmarks are not yet supported, because they have more than one initial state, include plant invariants or state requirement invariants, have no requirements, or have marker predicates in components.
To gauge the reduction in synthesis effort for the supported benchmarks, we compare the same BDD metrics as in Section~\ref{sec:experiments}, namely the number of BDD operations performed, and the maximum number of BDD nodes used.
For multilevel synthesis, we sum up the number of operations for the multiple syntheses for different parts of the system, and we take the maximum for the number of nodes.
The results are presented in Table~\ref{tbl:experiment-multilevel}.
In the third column, the number of groups in the multilevel system is shown.
Similar to Section~\ref{sec:experiments}, we again indicate a reduction factor.

For all benchmarks, the total number of operations and maximum number of nodes decrease or stay the same, when using multi-level synthesis.
For the MRI patient support system (event-based), there is a reduction of more than 42 times in the number of operations, and a reduction of 12 times in the maximum number of nodes.
For some of the other benchmarks there is also a significant reduction in effort, while for some the benefits are small.
Interestingly, for the production cell no reduction is achieved, even though the multi-level system consists of 11 groups.
For two benchmarks the multilevel system has only one group, and thus multilevel synthesis effectively amounts to monolithic synthesis.

To make efficient use of this technique, the plant components and requirements can first be split as much as possible, to avoid the emergence of large groups~\cite{GoordenMRFR19}.
Furthermore, `bus' components, that have connections to many other components, lead to clustering all these components and the associated requirements in the same group.
To address this challenge, `bus' components can be treated separately~\cite{GoordenDRMFR19}.
Applying these improvements to the current benchmarks is left as future work, since they are not yet implemented in Eclipse ESCET.
It is also future work to evaluate other non-monolithic synthesis approaches on our benchmarks.

\section{Conclusions and future work}
\label{sec:conclusions}

The Eclipse Supervisory Control Engineering Toolkit (ESCET) toolkit supports synthesis-based engineering (SBE) primarily through the CIF language and tools.
In this paper, we described CIF's symbolic synthesis algorithm, including aspects that are often omitted in the literature but are of great practical relevance, such as the prevention of runtime errors, handling different types of requirements, and supporting input variables.
We also introduced and described CIF's set of benchmark models, a collection of 23 industrial and academic models of various sizes and complexities, which are freely available.
Together, they allow researchers to more easily improve the synthesis algorithm and evaluate those improvements.

We also described recent improvements between ESCET versions v0.8 and v4.0 that affect synthesis performance, and evaluated them on our set of benchmark models.
This showed that synthesis performance has been significantly improved.
We then showed the current practical synthesis performance of CIF, which is sufficient in most cases.
However, for some models, still further improvements are needed.
We evaluated multi-level synthesis, a form of non-monolithic synthesis, showing that it can help reduce the synthesis effort.
But this, too, is not always sufficient.

As future work, we consider evaluation of multi-level synthesis with split requirements and better handling of `bus' components, as well as the evaluation of other non-monolithic synthesis approaches on our benchmarks.

\backmatter

\section*{Funding Declaration}

This research is partly carried out as part of the Poka Yoke program under the responsibility of TNO-ESI in cooperation with ASML and VDL-ETG.
Poka Yoke is funded by Holland High Tech -- TKI HTSM via the PPP Innovation scheme (PPP-I) for public-private partnerships (grant number TKI-HTSM/24.0352).

\section*{Data availability}

The artifact accompanying this paper includes the models and scripts that can be used to reproduce various results from this paper~\cite{ArtifactOfThisPaper}.
The artifact is available at Zenodo under identifier doi:10.5281/zenodo.13358811~\cite{ArtifactOfThisPaper}.
Eclipse ESCET v4.0, which is used for most of the experiments, can be downloaded at \url{https://eclipse.dev/escet/v4.0/download.html}.
Eclipse ESCET v0.8, against which we compare v4.0, can be downloaded at \url{https://eclipse.dev/escet/v0.8/download.html}.
The latest version of Eclipse ESCET can be downloaded at \url{https://eclipse.dev/escet/download.html}, but should not be used to reproduce the results from this paper.

\bibliography{main}

\end{document}